\documentclass[prb,preprint,superscriptaddress,onecolumn,10pt]{revtex4-1} 
\usepackage{amsmath}  % needed for \tfrac, \bmatrix, etc.
\usepackage{amsfonts} % needed for bold Greek, Fraktur, and blackboard bold
\usepackage{graphicx} % needed for figures
\usepackage{esvect}
\usepackage{mathtools}
\usepackage{subcaption}
\usepackage{xcolor}
\usepackage{appendix}
\usepackage{siunitx} 
\usepackage{placeins}

\usepackage{float} 

\usepackage{pbox}

\usepackage{mathtools}
\makeatletter
\newcommand{\vast}{\bBigg@{3}}
\newcommand{\Vast}{\bBigg@{3.5}}
\makeatother

\usepackage[mathscr]{euscript}
\DeclareSymbolFont{rsfs}{U}{rsfs}{m}{n}
\DeclareSymbolFontAlphabet{\mathscrsfs}{rsfs}

\newcommand{\dd}{\mathrm{d}}
\newcommand{\bd}{\mathscrsfs{D}}

\DeclareMathOperator{\expint}{E}

\begin{document}

% Be sure to use the \title, \author, \affiliation, and \abstract macros
% to format your title page.  Don't use lower-level macros to  manually
% adjust the fonts and centering.

\title{A Schr\"{o}dinger Equation for Evolutionary Dynamics}
% In a long title you can use \\ to force a line break at a certain location.

\author{Vi D. Ao}
\affiliation{Department of Physics, VNUHCM University of Science, 227 Nguyen Van Cu, Ho Chi Minh, 700000, Vietnam.}

\author{Duy V. Tran}
\thanks{These authors contributed equally to this work.}
\affiliation{Department of Mechanical Engineering, VNUHCM University of Technology, 226 Ly Thuong Kiet, Ho Chi Minh, 11000, Vietnam.}

\author{Kien T. Pham}
\thanks{These authors contributed equally to this work.}
\affiliation{Department of Aerospace Engineering, School of Transportation Engineering, Hanoi University of Science and Technology, 01 Dai Co Viet, Hanoi, 100000, Vietnam.}

\author{Duc M. Nguyen}
\thanks{These authors contributed equally to this work.}
\affiliation{University of Chicago, 5801 S Ellis Ave, Chicago, IL 60637, USA.}

\author{Huy D. Tran}
\affiliation{Department of Physics, The Hong Kong University of Science and Technology, Clear Water Bay, Kowloon, Hong Kong, 999077, PR China.}

\author{Tuan K. Do}
\affiliation{Department of Mathematics, UCLA, 520 Portola Plaza, Los Angeles, CA 90095-1555, USA.}

\author{Van H. Do}
\affiliation{Homer L. Dodge Department of Physics and Astronomy, Unversity of Oklahoma, 440 W. Brooks St. Norman, OK 73019, USA.}

\author{Trung V. Phan}
\email{trung.phan@yale.edu}
\affiliation{Department of Molecular, Cellular and Developmental Biology, Yale University, 260 Whitney Ave, New Haven, CT 06511, USA.}

\date{\today}

\begin{abstract}
We establish an analogy between the Fokker-Planck equation describing evolutionary landscape dynamics and the Schr\"{o}dinger equation which characterizes quantum mechanical particles, showing how a population with multiple genetic traits evolves analogously to a wavefunction under a multi-dimensional energy potential in imaginary time. Furthermore, we discover within this analogy that the stationary population distribution on the landscape corresponds exactly to the ground-state wavefunction. This mathematical equivalence grants entry to a wide range of analytical tools developed by the quantum mechanics community, such as the Rayleigh-Ritz variational method and the Rayleigh-Schr\"{o}dinger perturbation theory, allowing us to not only make reasonable quantitative assessments but also explore fundamental biological inquiries. We demonstrate the effectiveness of these tools by estimating the population success on landscapes where precise answers are elusive, and unveiling the ecological consequences of stress-induced mutagenesis -- a prevalent evolutionary mechanism in pathogenic and neoplastic systems. We show that, even in a unchanging environment, a sharp mutational burst resulting from stress can always be advantageous, while a gradual increase only enhances population size when the number of relevant evolving traits is limited. Our interdisciplinary approach offers novel insights, opening up new avenues for deeper understanding and predictive capability regarding the complex dynamics of evolving populations.
\end{abstract}

\maketitle % title page is now complete

\section{Introduction}

Evolution is the primary driving force behind the diversity and complexity of life on Earth for billions of years \cite{darwin2004origin,dawkins2016selfish}, allowing organisms to change and adapt over time \cite{waddington1959evolutionary,rose1996adaptation}. It emerges through the synergy between natural selection and genetic mutations, in which natural selection favors combinations of traits that enhance fitness \cite{endler1986natural} while genetic mutations introduce genetic variations that facilitate the emergence of new advantageous traits \cite{nevo1978genetic}. Within populations, ecological factors such as niche constraint \cite{tsoularis2002analysis,getz1991unified,getz1994metaphysiological} and environmental stress \cite{bjedov2003stress,fitzgerald2017stress} can exert their influence on these processes \cite{conrad1990geometry}, even molding the trajectory and tempo of evolution \cite{zhang2011acceleration}. 

The complex dynamics of evolving population can be captured by a Fokker-Planck equation on the evolutionary landscape \cite{risken1991fokker}, an abstract space of all possible genetic variations and their corresponding biological properties within a given ecological context \cite{wright1932roles}. The number of relevant evolving genetic traits corresponds to the dimensionality of this space \cite{vincent2005evolutionary}, where every combination corresponds to an unique position. On the landscape, together with the ecological influence, we represent the mutation process with an effective diffusion \cite{kimura1964diffusion} and the natural selection pressure with a fitness potential \cite{sergey2004fitness}. We show that there is an analogy between this formulation of evolutionary dynamics and the Schr\"{o}dinger description for quantum mechanical particles \cite{schrodinger1926quantisierung}, in which how a population evolves on the multi-dimensional landscape is almost similar to how a wavefunction behave under a multi-dimensional energy potential in imaginary time \cite{popov2005imaginary}. Following this observation, we further discover that the stationary population distribution on the landscape corresponds exactly to the quantum ground-state wavefunction \cite{schrodinger1926quantisierung,bloch1929quantenmechanik}. Such curious connection enables us to utilize various quantitative tools borrowed directly from quantum mechanics literature to quickly extract information about the steady population state of complex landscapes, even ones that lack exact analytical comprehension. For other examples of classical-quantum analogies where insights from quantum physics can illustrate classical phenomenon, see \cite{heidari2022tumor,fung2023analogy,armstrong2019variational,rozenman2023observation,rodrigues2022bright}.

Understanding the consequences of evolution has always been a cornerstone of biological research \cite{williams1970deducing}, serving as a fundamental pursuit aimed at unravelling the possible outcomes arising from this transformative force, and shedding light on the foundational principles that shape and govern all life on Earth. There exists many distinct evolutionary regimes \cite{mayr2001evolution}. For pathogenic and neoplastic populations such as microbial organisms and cancer cells, stress-induced mutagenesis \cite{bjedov2003stress,fitzgerald2017stress}, in which mutational increase can be triggered due to high biological stress, assumes a prominent role. Consider {\it E.coli} bacteria after experiencing exposure to antibiotics \cite{imasheva1999environmental,hoffmann2000environmental}, they undergo elongation and stop dividing \cite{phan2018emergence}. At the same time inside the bacterium, the SOS response switches on, leading to the induction of low-fidelity error-prone replication polymerases and, consequently, there is a sharp increase in the mutation rate during DNA replication from the typically low value of $D_l \propto 10^{-9}$ to a high rate of $D_h \propto 10^{-5}$ mutations per base pair per generation \cite{cirz2005inhibition, bos2015emergence}. There are also several other mechanisms by which genetic change can occur when organisms are under stress \cite{foster2007stress}. Laboratory studies have shown that at least $80\%$ of natural isolates of {\it E.coli} from diverse environments worldwide can exhibit stress-induced mutagenesis \cite{pribis2022stress}, highlighting its significance as an essential evolutionary dynamic in the realm of microbiology. Knowing the extensive effects of stress-induced mutagenesis in pathogenic and neoplastic systems is vital for developing strategies to combat their adaptive capabilities and improve therapeutic interventions \cite{govindaraj2018global}.

Here, thanks to the Schr\"{o}dinger analogy, we can conveniently employ two different methods: the Rayleigh-Ritz variational method\cite{rayleigh1907dynamical,ritz1909neue} to estimate the stationary population number, and the Rayleigh-Schr\"{o}dinger perturbation theory\cite{cohen1982rayleigh} to assess the tendency of population change resulting from stress-induced mutagenesis. For an unknown system evolving under a given Hamiltonian, the Rayleigh-Ritz variational method consists of finding trial wave functions that minimize the energy of the system to approximate the unknown ground state. Hence, we can estimate the stationary population size for a family of single-peak landscapes in all dimensions through this method. On the other hand, we also study the stress-induced mutagenesis using perturbation theory on a well known system: starting from a Hamiltonian with an well-known ground state and eigenenergy, we probe the evolution of this system perturbed by a small addition to this Hamiltonian using a power series expansion to the known ground state and the operator-representation of the additional Hamiltonian perturbation. This is the Rayleigh-Schrodinger  method, and we utilize it to consider two distinct extremes: gradual increases in mutation rate with stress and a sharp mutational burst when stress levels surpass a certain fitness threshold. We demonstrate in an unchanging environment that, unlike the former case, the latter consistently leads to a net gain in the total population size. This finding offers an explanation for the frequent appearance of mutational switches observed in nature \cite{cirz2005inhibition, bos2015emergence}.

\section{Evolutionary Landscape and Ecological Influence}

The landscape is typically represented as a multi-dimensional Euclidean space $\mathbb{R}^{\mathscrsfs{D}}$, where each point $\vec{x}$ represents a unique combination of $\mathscrsfs{D}$ scalar-strategy genetic traits \cite{vincent2005evolutionary}. When the maximum fitness $R(\vec{x})$ (which represents the selection pressure) remains constant over time \cite{wright1932roles,sergey2004fitness}, the population distribution density $b(\vec{x},t)$ within this landscape evolves via the Fokker-Planck equation \cite{risken1991fokker}:
\begin{equation}
\partial_t b = \nabla^2 \left( D b \right) + R b \ ,
\label{fokker_planck}
\end{equation}
where the effective diffusivity $D$ represents the local speed of mutations. In other words, higher value of $D$ results in a faster population diversification. 

\begin{figure*}[ht]
\centering
\includegraphics[width=0.8\textwidth,keepaspectratio]{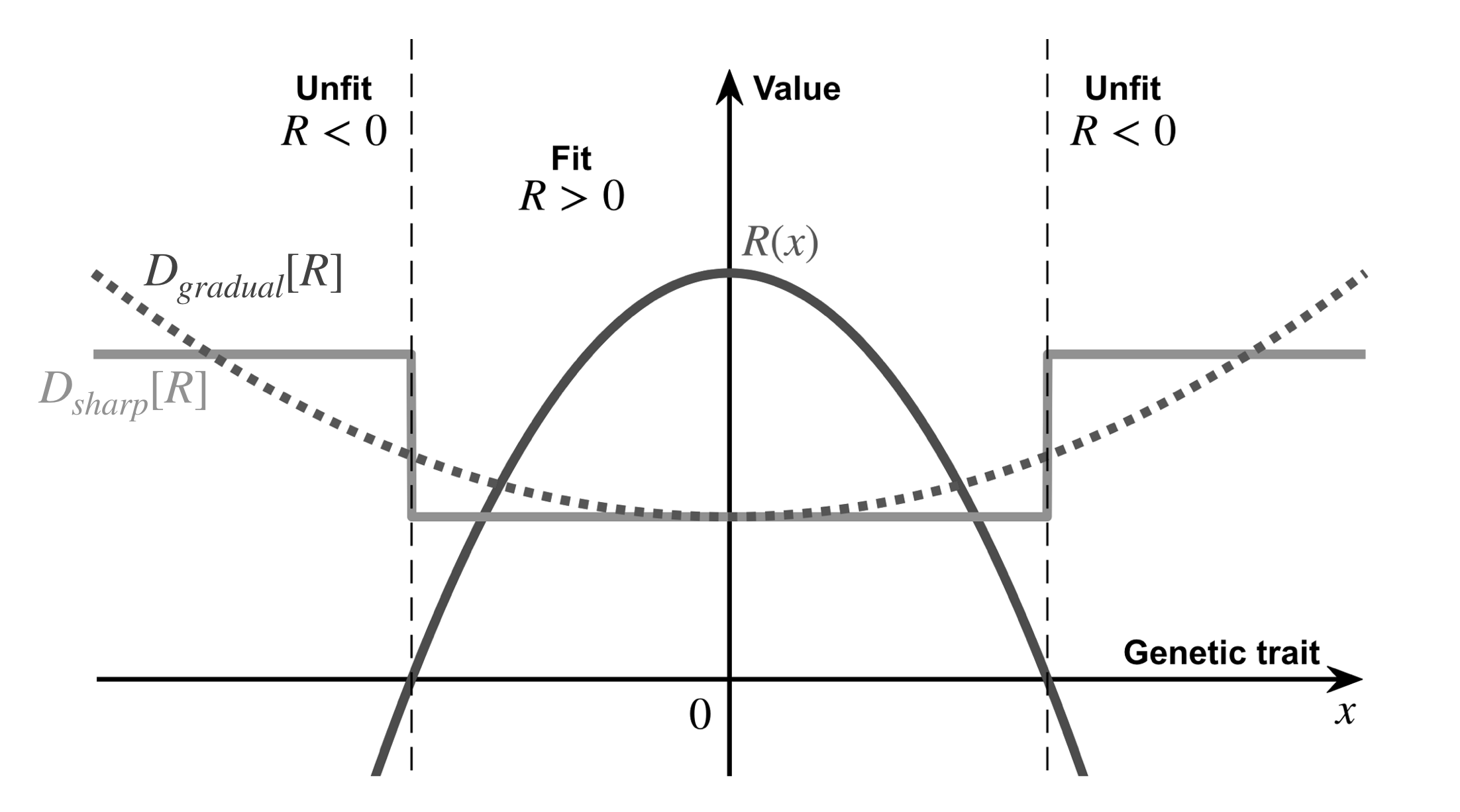}
\caption{\textbf{The evolutionary landscape of our stress-induced mutagenesis model, illustrated around a local peak fitness.} In this work, we consider two distinct regimes of stress-induced mutagenesis, corresponding to two different heterogeneous diffusion profiles: a gradual increased diffusivity $D_\text{gradual}[R]$ and a sharp transition diffusivity $D_\text{sharp}[R]$ at the transition from the fit to the unfit phenotype. In this cartoon we have used Eq. \eqref{toy_fitness} for $R(x)$, Eq. \eqref{gradual_diff} for $D_{\text{gradual}}[R]$, and Eq. \eqref{sharp_diff} for $D_{\text{sharp}}[R]$. Here $\max(R)=R(0)$, and for comparison between the two cases we use $D_{\text{gradual}}^{(0)} = D^{(0)}_{\text{sharp}}$.}
\label{fig1}
\end{figure*}

Our mathematical model is still incomplete as it assumes unlimited population growth. In any natural ecological system, the population growth of an organism should be limited by the resources available in its environment. The logistic model of population growth provides a better description of the population dynamics in a finite environment by taking into account the carrying capacity of the environment \cite{tsoularis2002analysis}. The carrying capacity $K$ in the logistic model of population growth \cite{getz1991unified,getz1994metaphysiological} represents the maximum number of individuals that can be sustained in a given environment. When the population size approaches the carrying capacity, the growth rate decreases until the population stabilizes at the carrying capacity. This carrying capacity can be incorporated into the mathematical framework by modifying Eq. \eqref{fokker_planck} into an integro-differential equation:
\begin{equation}
\partial_t b = \nabla^2 \left( D b \right) + \left[ 1- \frac{\displaystyle\int d^{\mathscrsfs{D}} \vec{x} b(\vec{x},t)}{K} \right] R b \ ,
\label{fokker_planck_logistic}
\end{equation}
in which the integration of population density distribution is the total population size $B(t)=\int d^{\mathscrsfs{D}} \vec{x} b(\vec{x},t)$. We can define the metric for population success as \cite{phan2021it}:
\begin{equation}
S(t) =\frac{B(t)}{K} = \frac{\displaystyle\int d^{\mathscrsfs{D}} \vec{x} b(\vec{x},t)}{K} \ ,
\label{success}
\end{equation}
then the expression for the growth rate is just $G = \left( 1-S \right) R$. This is a valid description at $S\leq 1$, and also exhibits a decrease of not only birth but also death rate at large population. For an example, bacteria such as {\it E.coli} can signal each others via quorum sensing, which can lead to a collective slowdown in metabolic rate at a dense bacterial population \cite{an2014bacterial}. In general, at high cell densities, the rate of cell death may decrease due to various reasons. One factor contributing to decreased cell death is the activation of stress responses and mechanisms that enhance cell survival. Bacteria can sense and respond to stressful conditions, such as nutrient limitation or high cell density, by activating protective mechanisms that increase cell viability and reduce cell death. This adaptive response can help bacteria survive and maintain population stability in crowded environments, which has been observed with bacteria living in biofilms \cite{mooney2018periprosthetic}. Rearrange the terms in Eq. \eqref{fokker_planck_logistic} and define a rescaled time $\tilde{t} = 2Dt$, we can arrive at:
\begin{equation}
-\partial_{\tilde{t}} b = \left( -\frac12 \nabla^2  - \frac{1-S}{2D} R \right) b \ ,
\label{fokker_planck_logistic_rearrange}
\end{equation}
which has the form of a hyperbolic differential equation if the success is treated as a constant.

 In order to describe stress-induced mutagenesis, the effective diffusivity $D$ should not be a constant. This evolutionary regime is a major concern for medical research due to its ability to accelerate the development of drug resistance in pathogenic and neoplastic systems \cite{zhang2011acceleration,wu2014game,li2021acceleration}, while also creating other complications in the treatment of infectious diseases \cite{ram2014stress}. According to the World Health Organization, antibiotic-resistant infections caused an estimated $1.27$ million deaths worldwide in 2019 \cite{murray2022global}. Stress-induced mutagenesis has also been found to have a notable impact on the evolution of the SARS-CoV-2 virus and the emergence of novel variants \cite{kemp2021sars}, highlighting the need for better understanding. One of the most crucial biological inquiries one could ask about stress-induced mutagenesis is why it behaves in the way it does. There are many possible functional-dependence of mutation rate on stress, yet nature somehow seems favor a mutational switch \cite{cirz2005inhibition,bos2015emergence}. To explore this further, we cast this question into the mathematical framework presented by Eq. \eqref{fokker_planck_logistic_rearrange}. To capture the intricacies of stress-induced mutagenesis, we incorporate a heterogeneous effective diffusivity into the landscape. We seek to investigate the theoretical distinctions between the outcomes of these two different possibilities for the diffusivity $D$  as a function of fitness, which is defined as the ability to reproduce. In the gradual case, a linearity governs:
\begin{equation}
D_\text{gradual}[R] = D^{(0)}_\text{gradual} + D^{(1)}_\text{gradual} \Delta R \ ,
\label{gradual_diff}
\end{equation}
in which $\Delta R = \max (R) - R$ is the difference between the maximum fitness $\max(R)$ and the fitness $R$, whereas the sharp case follows a Heaviside step-function in which the transition happens right at the boundary between the fit and the unfit regions:
\begin{equation}
D_\text{sharp}[R] = D^{(0)}_\text{sharp} + D^{(1)}_\text{sharp} \Theta(-R) \ .
\label{sharp_diff}
\end{equation}
$\Theta(\zeta)$ is the Heaviside function, in which $\Theta(\zeta<0)=0$ and $\Theta(\zeta>0)=1$. Fig. \ref{fig1} serves as a visual representation of the basic postulations underlying our theoretical analysis. We summary the biophysical quantities used in our mathematical model in Appendix \ref{summary_math}.

\section{An Analogy to the Schr\"{o}dinger Equation} 

The analogy between the Fokker-Planck equation as in Eq. \eqref{fokker_planck_logistic_rearrange} and the Schr\"{o}dinger equation \cite{schrodinger1926quantisierung,bloch1929quantenmechanik} can be elucidated by considering the following identifications. We introduce an imaginary time variable $i\tau$ related to the diffusion coefficient $D$ and the physical time $t$ and an energy potential $V(S,\vec{x})$ related to the population success $S$ maximum growth rate $R(\vec{x})$: 
\begin{equation}
i\tau \leftrightarrow \tilde{t} \ , \ V(S,\vec{x}) \leftrightarrow -\frac{1-S}{2D} R(\vec{x}) \ .
\label{identification}
\end{equation}
This procedure of changing from real time to imaginary time is known as Wick rotation \cite{kontsevich2021wick}. If we treat $S$ as a constant parameter, then for Planck constant $\hbar=1$ and mass $m=1$, we can recast Eq. \eqref{fokker_planck_logistic_rearrange} into a form that closely resembles the Schr\"{o}dinger equation in imaginary time $\tau$:
\begin{equation}
i\hbar \partial_\tau \Psi = \hat{H} \Psi \ , \ \hat{H} = \frac{\hat{p}^2}{2m} + V(S,\vec{x}) \ , 
\label{Schrodinger_Bloch_imaginary_time}
\end{equation}
where $\hat{p} = - i \hbar \nabla$ is the momentum operator and $\Psi(\vec{x},t) \propto b(\vec{x},t)$ represents the wavefunction of a single quantum mechanical particle of mass $m$ moving in our multi-dimensional landscape. The Hamiltonian operator $\hat{H}$ governs the behavior of this particle under the influence of the energy potential $V(S,\vec{x})$. From here on, we drop $\hbar$ and $m$ out of the analysis.

While this analogy is not exact, as it assumes that success $S$ is unchanging, independent of distribution density $b(\vec{x},t)$, and therefore neglects the influence of the total population number on the potential energy function, it can still provide a powerful framework for understanding the dynamics of evolution. We demonstrate that by looking at the stationary state, where $b=b_\text{st}(\vec{x})$ is an unchanging spatial-function and thus $S=S_\text{st}$ is fixed. We now have an exact correspondence between Eq. \eqref{Schrodinger_Bloch_imaginary_time} and a time-independent Schr\"{o}dinger equation associated with $E=0$ eigenstate (also known as the stationary Schr\"{o}dinger equation): 
\begin{equation} 
0 = \left( -\frac12 \nabla^2  - \frac{1-S_\text{st}}{2D} R \right) b_\text{st} \ \longleftrightarrow \ E \Psi_\text{st} = \left( \frac{\hat{p}^2}{2m} + V(S_\text{st},\vec{x}) \right) \Psi_\text{st} \ .
\label{popdyn_to_Schro}
\end{equation}
Here we obtain a powerful constraint -- the stationary population success $S_\text{st}$ must correspond to an energy potential $V(S_\text{st},\vec{x})$ that has a zero-eigenenergy. Moreover, since the population density $b_\text{st}(\vec{x})$ is a non-negative physical field, its associated wavefunction $\Psi_\text{st}(\vec{x})$ should not change sign and cross zero anywhere on the entire landscape (one exception is at impenetrable boundaries, where the wave function is forced to vanish) \cite{van2007fundamental}. This further restriction implies that the wavefunction should also be the ground state of the energy potential $V(S_\text{st},\vec{x})$. Together, we require $V(S_\text{st},\vec{x})$ to be a potential energy function that possesses a ground-state with zero-energy, and $\Psi(\vec{x})$ must be the ground-state wavefunction $\Psi_{\Omega}(\vec{x})$.

This kind of quantum mechanical analogy in classical biological phenomena has been discovered in other contexts as well. For instance, the very same Schr\"{o}dinger equation we consider in our paper also emerges in bacterial chemotaxis \cite{rosen1983theoretical} as well, although only at a very special subset -- but experimentally has been observed in actual bacteria populations -- of the parameter space.

\begin{figure*}[ht]
\centering
\includegraphics[width=\textwidth,keepaspectratio]{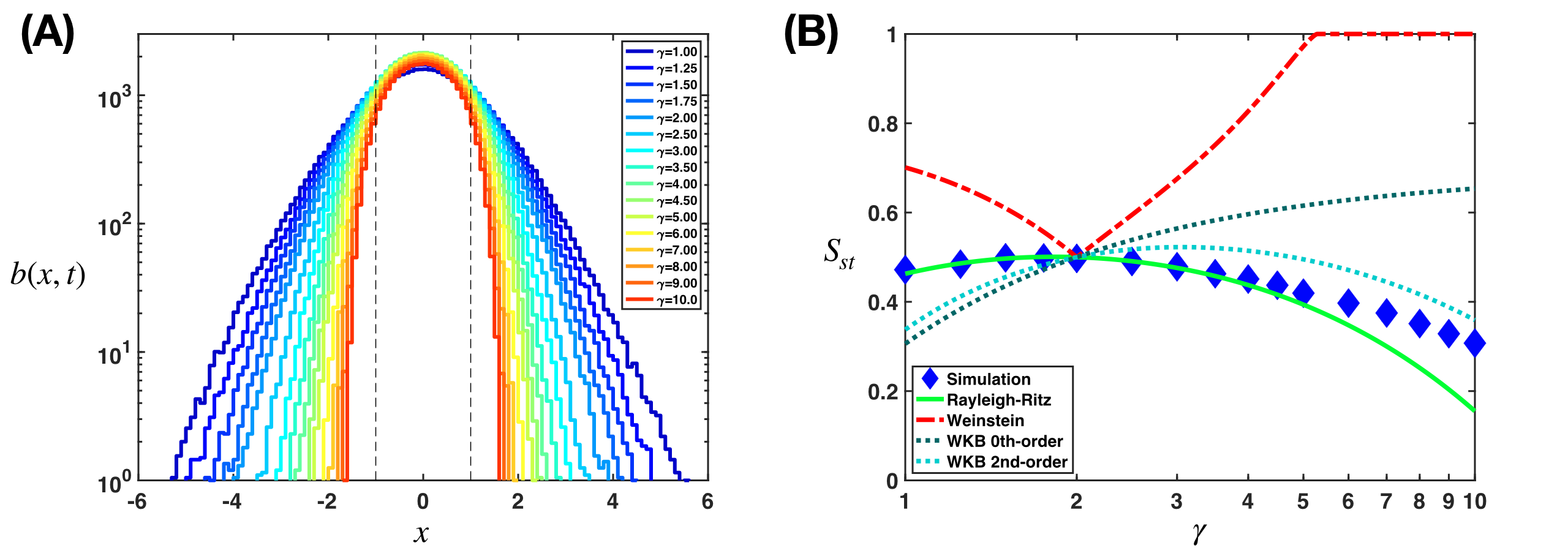}
\caption{\textbf{Stationary population distributions and successes for different fitness landscapes can be estimated with the Rayleigh-Ritz variational method.} (A) The distributions of the heterogeneous population on different fitness landscapes at the stationary state, where the fitness obeys $R(\vec{x})=R_0 \left[ 1 - \left( \frac{|\vec{x}|}{\lambda} \right)^\gamma \right]$ and the exponent $\gamma \in [1,10]$, concentrates more around the optimal position $x_{\text{op}}=0$ with increasing $\gamma$. The dash-lines mark $x=\pm \lambda$, where the fitness hits $0$. Here we show the results from a simulation, which we described in Appendix \ref{simulation_explain}. Here we consider $\bd=1$-dimensional landscapes and use the parameter values $D=1/2$, $R_0=1$, $\lambda=1$, and $K=10^5$. (B) We compare different methods of estimation for the stationary population success with the simulation findings. We show the analytical results as obtained from the Rayleigh-Ritz variational method as in Eq. \eqref{Sst_RR_main}, the Weinstein method as in Eq. \eqref{Sst_W}, and the Wentzel-Krammers-Brillouin approximation (0th- and 2nd-order) as in Eq. \eqref{Sst_WKB0} and Eq. \eqref{Sst_WKB2}.}
\label{fig2}
\end{figure*}

Let us show how to utilize this convenient constraint in practice. Consider an inverse-quadratic maximum growth rate $R(x)$ peaked and centered around the optimal combinations of genetic traits which is chosen to be at $\vec{x}_{\text{op}} = 0$:
\begin{equation}
R(\vec{x}) = R_0 \left[ 1 - \left( \frac{\vert \vec{x} \vert}{\lambda} \right)^2 \right]
\label{toy_fitness} \ ,
\end{equation}
It means the further away from the origin, the less fit an organism becomes. We call $\vert \vec{x} \vert <\lambda$ the {\it fit region} where $R>0$, and $\vert \vec{x} \vert >\lambda$ the {\it unfit region} where $R<0$ \cite{phan2021it}, as already shown in Fig. \ref{fig1}. Following Eq. \eqref{identification}, this fitness landscape corresponds to a simple harmonic oscillator potential energy $U_{\text{SHO}}(\vec{x})$ up to a shift $U_0$:
\begin{equation}
V(S_\text{st},\vec{x}) = U_{\text{SHO}} (\vec{x}) + U_0 
 \ , \ U_{\text{SHO}} (\vec{x}) = \frac12 \omega^2 \vert \vec{x} \vert^2 \ ,
 \label{SHO_and_shift}
\end{equation}
in which the angular oscillation frequency and the downward shift are:
\begin{equation}
\omega^2 = \frac{1-S_\text{st}}{D\lambda^2} R_0 \ , \ U_0 = -\frac12 \omega^2 \lambda^2 \ .
\label{freq_and_shift}
\end{equation}
The ground-state energy of a $\mathscrsfs{D}$-dimensional oscillator, which corresponding to the purely-quadratic potential $U_{\text{SHO}}(\vec{x})$, is a fundamental result that can be found in pretty much every quantum mechanics textbooks \cite{dirac2001lectures,landau2013quantum,griffiths2018introduction,sakurai1995modern}): 
\begin{equation}
E_{\Omega} = \mathscrsfs{D} \frac{\omega}2 \ .
\label{SHO_groundstate_energy}
\end{equation}
So for it to be $0$ after the energy shift, we need:
\begin{equation}
E = E_{\Omega} + U_0 = 0 \ \Longrightarrow \ \omega = \frac{\mathscrsfs{D}}{\lambda^2} \ , 
\label{set_to_0}
\end{equation}
which directly gives us the stationary population success from Eq. \eqref{freq_and_shift}:
\begin{equation}
S_\text{st} = 1 - \frac{D\lambda^2}{R_0}\omega^2 = 1 - \frac{\mathscrsfs{D}^2 D}{R_0 \lambda^2} \ .
\label{pop_suc}
\end{equation}
If $S_\text{st}<0$, it means there is no sustainable success, and the population eventually goes extinct on such ecological system. We get the stationary population distribution density $b_\text{st}(\vec{x})$, starting from the Gaussian ground-state wavefunction of a simple harmonic oscillator \cite{dirac2001lectures,landau2013quantum,griffiths2018introduction,sakurai1995modern}:
\begin{equation}
b_\text{st}(\vec{x}) \propto \Psi_{\Omega}(\vec{x}) \propto \exp\left( -\frac12 \omega \vert\vec{x}\vert^2 \right) = \exp\left[ -\frac{\mathscrsfs{D}}2 \left( \frac{\vert\vec{x}\vert}{\lambda} \right)^2 \right] \ .
\label{bst_from_PsiOmega}
\end{equation}
Using Eq. \eqref{success}, we can determine the pre-factor and obtain:
\begin{equation}
b_\text{st}(\vec{x}) = \frac{K }{\sqrt{2\pi \lambda^2 / \mathscrsfs{D}}^{\mathscrsfs{D}}} \left( 1 - \frac{\mathscrsfs{D}^2 D}{R_0 \lambda^2}\right)\exp\left[ -\frac{\mathscrsfs{D}}2 \left( \frac{\vert\vec{x}\vert}{\lambda} \right)^2 \right] \ .
\label{bst_SHO}
\end{equation}
Similar analytical investigations can be done to extract $S_\text{st}$ and $b_\text{st}(\vec{x})$ from the quantum mechanical ground-state, for any function $R(\vec{x})$ defined on the landscape. 

We can utilize the Rayleigh-Ritz variational method \cite{rayleigh1907dynamical,ritz1909neue} to estimate the upper-bound (and also the Weinstein method \cite{weinstein1934modified,lee1987upper} for the lower-bound) of the population success $S_\text{st}$, in any dimensions. As a demonstration, we consider a class of landscapes with power-law dependency fitness  $R(\vec{x})=R_0 \left[ 1- (|\vec{x}|/\lambda)^\gamma \right]$. The fitness we considered before, given by Eq. \eqref{toy_fitness}, belongs to this class and corresponds to the exponent value $\gamma = 2$. For a general value of $\mathscrsfs{D}$ and $\gamma$, the exact solution for the ground-state is not known. But using a Gaussian ansatz-wavefunction we can quickly estimate the stationary population success $S_{\text{st}}$ and the width $\sigma$ of the population distribution around the optimal peak on the landscape:
\begin{equation}
S_{\text{st}} \geq S^{\text{(RR)}}_{\text{st}}= 1 - \frac{2D}{R_0 \lambda^2}  (2+\gamma)^{\frac{2+\gamma}{\gamma}} \left( \frac{\mathscrsfs{D}}{4\gamma} \right) \left[ \frac{\Gamma\left( \frac{\mathscrsfs{D}+\gamma}2\right)}{2\Gamma\left( \frac{\mathscrsfs{D}}2\right)}  \right]^{\frac{2}{\gamma}} \ , \ \sigma \approx \left[ \frac{\mathscrsfs{D}}{2\gamma} \frac{\Gamma\left( \frac{\mathscrsfs{D}}2\right)}{\Gamma\left( \frac{\mathscrsfs{D}+\gamma}2\right)} \right]^{\frac1{2+\gamma}} \ .
\label{Sst_RR_main}
\end{equation}
We will carry out in details this estimation with a Gaussian trial-wavefunction in Appendix \ref{rayleigh_ritz}. The width $\sigma$ decreases with the exponent $\gamma$, which is consistent with our simulation findings as shown in Fig. \ref{fig1}A. There are also other methods for estimating the ground state, such as the lesser known Weinstein method \cite{weinstein1934modified} and the Temple method \cite{temple1928theory,pollak2019tight} which can give us lower-bounds, and the one-dimensional Wentzel-Kramers-Brillouin approximation \cite{wentzel1926verallgemeinerung,kramers1926wellenmechanik,brillouin1926mecanique,karnakov2012wkb,voros1983return,balian1978discrepancies} which does not admit a simple higher-dimensional generalization but has gained more interests recently via the exact quantization condition \cite{gabai2023exact,ito2019tba,dorey1999anharmonic,voros1999airy}. We will go through the Weinstein method and the Wentzel-Kramers-Brillouin approximation in Appendix \ref{weinstein} and Appendix \ref{wkb}. In Fig. \ref{fig2}B, we compare the estimations for $S_{\text{st}}$ using different methods with results from simulation. We describe our simulation in Appendix \ref{simulation_explain}. 

In Fig. \ref{fig3} we show the analytical results obtain from the estimations of $S_{\text{st}}$, with the Rayleigh-Ritz variational method as in Eq. \eqref{Sst_RR_main}, the Weinstein method as in Eq. \eqref{Sst_W}, and the Wentzel-Krammers-Brillouin approximation (0th- and 2nd-order) as in Eq. \eqref{Sst_WKB0} and Eq. \eqref{Sst_WKB2}. Rayleigh-Ritz variation method, despite its simplicity, captures correctly and consistently the non-monotonic behavior of $S_{\text{st}}(\gamma)$ at small exponent $\gamma$, which peaks at $\gamma \approx 1.83$ and deviations from the simulation values for the region $\gamma\in [1,4]$ are less that $4\%$. Biophysically, small values of $\gamma$ can be interpreted as corresponding to weakly-curved fitness landscapes where fitness do not drastically decrease with mutations that go away from the optimal position. Weinstein method, which follows naturally from Rayleigh-Ritz variational method, requires much more calculations, and in general gives a bad estimation (except for around $\gamma\rightarrow 2$). We also explain the sudden change at $\gamma=2$ of its estimated $S_{\text{st}}$ in Appendix \ref{weinstein}. The Wentzel-Krammers-Brillouin approximation is very off at 0th-order (except also for $\gamma \rightarrow 2$) but can get much better at 2nd-order. We note that this inability to describe the ground state at high precision is a feature expected from this approximation, which works best at highly excited states where the wavelengths are much smaller than that of the potential characteristic length-scale \cite{vranivcar2000accuracy}.

\section{Applying Rayleigh-Schr\"{o}dinger Perturbation Theory to Stress-Induced Mutagenesis}

We investigate the impact of stress-induced mutagenesis on the stationary population size $B_\text{st}$ in both cases, as listed in Eq. \eqref{gradual_diff} and Eq. \eqref{sharp_diff}, for all natural dimensionality $\mathscrsfs{D} \in \mathbb{N}$ of the landscape. Rather than attempting to solve the exact, non-tractable evolution dynamics on the landscape, we adopt a perturbative approach. This enables us to reveal distinctions between the two cases in a tractable manner. We split the Hamiltonian in Eq. \eqref{Schrodinger_Bloch_imaginary_time} into the unperturbed $\hat{H}_0$ and the perturbed $\epsilon \hat{H}_p$. At the stationary state:
\begin{equation}
\hat{H} = \hat{H}_0 + \epsilon \hat{H}_p \ , \ \hat{H}_0 = \frac12 \hat{p}^2 + V(S_\text{st},\vec{x}) \ ,
\label{unperturbed_perturbed}
\end{equation}
where the energy potential $V(S_\text{st},\vec{x})$ is as given in Eq. \eqref{SHO_and_shift} and Eq. \eqref{freq_and_shift}. At the lowest-order of perturbation $\mathscrsfs{O}(\epsilon)$, the correction $\epsilon \delta E_{\Omega}$ to the ground-state energy in Eq. \eqref{SHO_groundstate_energy} can be estimated via Rayleigh-Schr\"{o}dinger perturbation theory even for a non-Hermittian $\hat{H}_p$ \cite{cohen1982rayleigh}: 
\begin{equation}
E_\Omega = \mathscrsfs{D} \frac{\omega}2 + \epsilon \delta E^{(1)}_{\Omega} \ , \ \delta E^{(1)}_{\Omega} = \frac{\displaystyle\int d^{\mathscrsfs{D}}\vec{x} \  \Psi_\Omega^{\dagger}(\vec{x}).\hat{H}_p. \Psi_\Omega(\vec{x})}{\displaystyle\int d^{\mathscrsfs{D}}\vec{x} \  \Psi_\Omega^{\dagger}(\vec{x}) \Psi_\Omega(\vec{x})} \ ,
\label{perturbative_correction}
\end{equation}
where $\Psi_\Omega(\vec{x})$ is the ground-state wavefunction of the unperturbed Hamiltonian as given by Eq. \eqref{bst_from_PsiOmega}. Our analysis can unveil the tendencies with which different manifestations of stress-induced mutagenesis affect the population, either boosting or suppressing success $S_\text{st}$.

\subsection{A Gradual Change}

The diffusivity on the landscape as in Eq. \eqref{gradual_diff}, which is associated with a gradual change in mutation rates, can be expressed as followed: 
\begin{equation}
D_\text{gradual} = D\left[1 + \epsilon \left( \frac{\vert \vec{x}\vert}{\lambda}\right)^2 \right] \ . \ 
\label{gradual_diff_rewrite}
\end{equation}
where we define the constants $D$ and $\epsilon$ to be:
\begin{equation}
D=D^{(0)}_\text{gradual} \ , \ \epsilon = \frac{D^{(1)}_\text{gradual} R_0}{D^{(0)}_\text{gradual}} \ . \ 
\end{equation}
We consider $\epsilon$ as a perturbation parameter, in the limit $\epsilon \ll 1$ which corresponds to $D^{(0)}_\text{gradual} \gg D^{(1)}_\text{gradual} R_0$.

The perturbed Hamiltonian in Eq. \eqref{unperturbed_perturbed} for this regime of stress-induced mutagenesis is given by a non-Hermitian operator:
\begin{equation}
\hat{H}_p = \frac12 \hat{p}^2 \left( \frac{\vert \vec{x} \vert}{\lambda} \right) ^2 \ .
\label{gradual_perturb_Hamiltonian}
\end{equation}
Applying Eq. \eqref{perturbative_correction}, we obtain:
\begin{equation}
\delta E^{(1)}_{\Omega} = \frac{1}{8\lambda^2} \mathscrsfs{D}\left( \mathscrsfs{D}-2\right) \ .
\end{equation}
The detail of this calculation can be found in Appendix \ref{perturbative_calc_1}.

Applying Eq. \eqref{set_to_0} including the ground-state energy correction:
\begin{equation}
E = E_\Omega + U_0 = \mathscrsfs{D} \frac{\omega}2 + \epsilon \frac{1}{8\lambda^2} \mathscrsfs{D}\left( \mathscrsfs{D}-2\right) - \frac12 \omega^2 \lambda^2 = 0\ ,
\end{equation}
we can approximate $\omega$ at the first-order of $\epsilon$-expansion:
\begin{equation}
\omega \approx \frac{\mathscrsfs{D}}{\lambda^2} \left[ 1 + \epsilon \eta(\mathscrsfs{D}) \right] \ , \ \eta(\mathscrsfs{D}) = \frac{\mathscrsfs{D} - 2}{4\mathscrsfs{D}} \ .
\label{eta_gradual}
\end{equation}
For a high dimensionality $\mathscrsfs{D} > 2$, $\eta(\mathscrsfs{D})$ is a positive value. Perturbative stress-induced mutagenesis in this regime increases $\omega \uparrow$. Since Eq. 
\eqref{freq_and_shift} indicates that $\omega$ and $S_{\text{st}}$ have  an inverse monotonic relationship, this means we get a reduction in success $S_{\text{st}} \downarrow$. In other words, stress-induced mutagenesis tends to suppress the population success when the number of relevant genetic traits on the landscape is high.

The above statement does not change if we consider another power-law dependency for the perturbative Hamiltonian in Eq. \eqref{gradual_perturb_Hamiltonian}, since:
\begin{equation}
\hat{H}_p \propto \hat{p}^2 \vert \vec{x} \vert^\kappa \ \ \Longrightarrow \ \ \eta (\mathscrsfs{D}) \propto \mathscrsfs{D} - \kappa \ . 
\end{equation}
We get $\eta (\mathscrsfs{D}) > 0$ when $\mathscrsfs{D} > \kappa$. The derivation can be found in Appendix \ref{perturbative_calc_2}.

\subsection{A Sharp Change}

We can rewrite Eq. \eqref{sharp_diff}, which described the diffusivity on the landscape associated with a sharp change in mutation rates, as follows:
\begin{equation}
D_\text{sharp} = D\left[1 + \epsilon \Theta \left( \frac{\vert \vec{x} \vert}{\lambda} - 1 \right) \right] \ . \ 
\label{sharp_diff_rewrite}
\end{equation}
From Eq. \eqref{toy_fitness} and Eq. \eqref{sharp_diff}, we define the constants $D$ and $\epsilon$ to be:
\begin{equation}
D=D^{(0)}_\text{sharp} \ , \ \epsilon = \frac{D^{(1)}_\text{sharp}}{D^{(0)}_\text{sharp}} \ . \ 
\end{equation}
Here we treat the up-step contribution as perturbation $\epsilon \ll 1$, which requires $D^{(0)}_\text{sharp} \gg D^{(1)}_\text{sharp}$, for subsequent calculations to be analytically tractable.

For this posibility of stress-induced mutagenesis, the perturbed Hamiltonian in Eq. \eqref{unperturbed_perturbed} is given by the following non-Hermitian operator:
\begin{equation}
\hat{H}_p = \frac12 \hat{p}^2 \Theta \left( \frac{\vert \vec{x} \vert}{\lambda} - 1 \right) \ ,
\label{Sharp_perturb_Hamiltonian}
\end{equation}
Following Eq. \eqref{perturbative_correction}, we can make the estimation:
\begin{equation}
\delta E^{(1)}_{\Omega} = \frac1{\lambda^2} \frac{(\omega\lambda^2)^{\frac{\mathscrsfs{D}}2 +1} \left[ -2e^{-\omega\lambda^2} + \omega \lambda^2\expint_{-\frac{\mathscrsfs{D}}2} \left( \omega\lambda^2 \right) \right]}{2\Gamma\left( \frac{\mathscrsfs{D}}2 \right)} \ ,
\end{equation}
where we remind that $\omega$ is as given in Eq. \eqref{freq_and_shift}. We carry out the details of this calculation in Appendix \ref{perturbative_calc_3}. We can then determine $\omega$ using Eq. \eqref{set_to_0}:
\begin{equation}
\begin{split}
E & = E_{\Omega} + U_0  
\\
& = \mathscrsfs{D} \frac{\omega}2 + \epsilon \frac1{\lambda^2} \frac{(\omega\lambda^2)^{\frac{\mathscrsfs{D}}2 +1} \left[ -2e^{-\omega\lambda^2} + \omega \lambda^2\expint_{-\frac{\mathscrsfs{D}}2} \left( \omega\lambda^2 \right) \right]}{2\Gamma\left( \frac{\mathscrsfs{D}}2 \right)}   - \frac12 \omega^2 \lambda^2 = 0 \ , 
\end{split}
\end{equation}
in which we obtain the approximate solution:
\begin{equation}
\omega \approx \frac{\mathscrsfs{D}}{\lambda^2} \left[ 1 + \epsilon \eta(\mathscrsfs{D}) \right] \ , \ \eta(\mathscrsfs{D}) =\frac{\mathscrsfs{D}^{\frac{\mathscrsfs{D}}2-1} \left[ -2e^{-\mathscrsfs{D}} + \mathscrsfs{D} \expint_{-\frac{\mathscrsfs{D}}2} \left( \mathscrsfs{D} \right) \right]}{\Gamma\left( \frac{\mathscrsfs{D}}2 \right)}  \ ,
\label{eta_sharp}
\end{equation}
where $\expint_{\ldots}(\ldots)$ is the generalized exponential integral \cite{olver1994generalized}. Since $\eta(\mathscrsfs{D})<0$ for every natural dimensionality $\mathscrsfs{D} \in \mathbb{N}$, this perturbative stress-induced mutagenesis effect always reduces $\omega \downarrow$, thus increases success $S_{\text{st}} \uparrow$ as follows from Eq. \eqref{freq_and_shift}.

The nonperturbative stationary solution of this case has been studied in our previous work \cite{2303.09084}. For a sanity check, one can show that what we have found here using Rayleigh-Schr\"{o}dinger perturbation theory agrees with the exact result at the leading-order of the perturbative parameters $\epsilon$.

\begin{figure*}[ht]
\centering
\includegraphics[width=\textwidth,keepaspectratio]{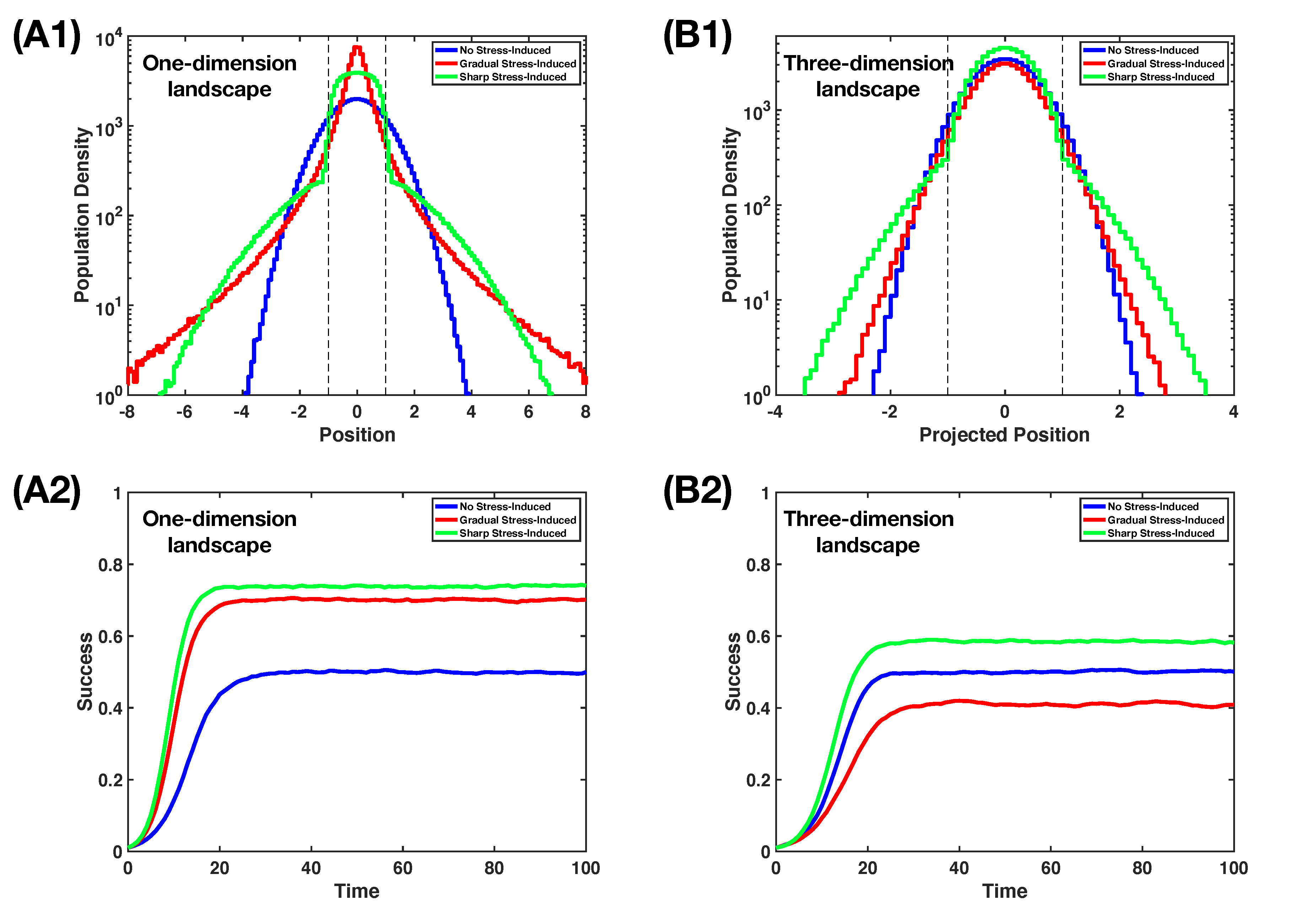}
\caption{\textbf{Rayleigh-Schrodinger perturbation theory can predict how different regimes of stress-induced mutagenesis affect the population success, extrapolatable beyond the perturbative regime.} Here we show our simulation findings, which we use the parameter values $R_0=1$, $
\lambda=1$, $K=10^5$, and fitness as in Eq. \eqref{toy_fitness}. We describe our simulation in Appendix \ref{simulation_explain}. (A1) The population distributions at stationary state on $\bd=1$-dimensional landscape for no stress-induced, gradual stress-induced, and sharp stress-induced mutagenesis regimes. The dash-lines mark $x=\pm \lambda$. For gradual stress-induced regime we use Eq. \eqref{gradual_diff_rewrite} with $\epsilon=10$, for sharp stress-induced regime we use Eq. \eqref{sharp_diff_rewrite} with $\epsilon=10$. (A2) The evolution of population success $S(t)$ with time $t$ on $\bd=1$-dimensional landscape for different mutagenesis regimes. (B1) The population distributions at stationary state on $\bd=3$-dimensional landscape. We consider looking at the distribution from a projection, i.e. $\vec{x}=(x_1,x_2,x_3)$ then we can use $x_1$ as the projected position. The dash-lines mark $x_1=\pm \lambda$. For gradual stress-induced regime we use Eq. \eqref{gradual_diff_rewrite} with $\epsilon=0.1$. Note that with $\epsilon=10$, the population goes extinct, as $S_{\text{st}}=0$. For sharp stress-induced regime we use Eq. \eqref{sharp_diff_rewrite} with $\epsilon=10$. (B2) The evolution of population success $S(t)$ with time $t$ on $\bd=3$-dimensional landscape.}
\label{fig3}
\end{figure*}

\subsection{A Comparison between two Stress-Induced Mutagenesis Regimes}

While the calculations in the previous section are done with perturbation theory, which corresponds to weak stress-induced mutagenesis effects, our findings go beyond that, as shown with simulations in Fig. \ref{fig3} for large values of $\epsilon$. We describe our simulation in Appendix \ref{simulation_explain}. Gradual stress-induced mutagenesis, which can be beneficial on low-dimensional landscapes (see Fig. \ref{fig3}A1 and Fig. \ref{fig3}A2), becomes quite lethal at high-dimensional landscapes (see Fig. \ref{fig3}B1 and Fig. \ref{fig3}B2). Sharp stress-induced mutagenesis, on the other hand, always up the stationary population success.

To get some intuitive understanding of how these two regimes can be so different, consider the followings. The expression for the total diffusive flux of population on the landscape is given by $J = \nabla(Db)$, which can be further broken down into two contributions: the diffusion-gradient contribution $J_\text{diff} = (\nabla D) b$ which is driven by the local slope of the diffusivity, and the density-gradient contribution $J_\text{dens} = D(\nabla b)$ which is generated by the heterogeneity of population distribution. In the case of a gradually changing $D_\text{gradual}[R]$, the contribution $J_\text{diff}$  exists generally everywhere, pointing towards the optimal combination of genetic traits. This means there is a clear guidance toward peak fitness on the $\mathscrsfs{D}$-dimensional space of genetic variations, focusing the population into a specific hypervolume. In contrast, for a sharp transition $D_\text{sharp}[R]$, this flux vanishes everywhere except at the $(\mathscrsfs{D}-1)$-dimensional boundary hypersurface between fit and unfit regions. Thus, intuitively, we expect that abruptly increasing mutation rates via stress-induced mutagenesis may be less effective in maintaining large stationary population size.

Our application of Rayleigh-Schr\"{o}dinger perturbation theory has quickly revealed a paradoxical outcome that influence evolutionary dynamics in the presence of multiple relevant evolving genetic traits. We interpret this mathematical finding as follows: in the case of gradual stress-induced mutagenesis, the diffusive-gradient flux becomes less effective at population concentration towards the fit region, as it spreads out excessively as the number of landscape dimensions increases. Conversely, sharp stress-induced mutagenesis enables the diffusive-gradient flux to remain spatially focused, even singular, and thus boosts the population success consistently, regardless of the number of genetic traits involved. It has been obsered that natural selection favors species that can optimize multiple biological capabilities simultaneously \cite{endler1986natural,arnold1983morphology}. As a result, the sharp regime of stress-induced mutagenesis may be preferred. Empirical evidence supports this notion \cite{cirz2005inhibition, bos2015emergence}. Here we have shown a quantitative argument for why this might be the case.

\section{Discussion} 

In this study, we reveal a curious analogy between a fundamental equation in quantum mechanics and the equation that governs the evolution of multiple genetic traits in a population. We show that determining the stationary distribution of a heterogeneous population can be mapped to the problem of finding the ground state wavefunction, fostering a more unified understanding of diverse phenomena and leading to new analytical approaches. Techniques developed for dealing with quantum mechanical systems can be adapted and applied to comprehend population dynamics, not only more expeditiously but also more profoundly.

The Rayleight-Ritz variational method \cite{rayleigh1907dynamical,ritz1909neue} allows us to quickly estimate the population number at equilibrium state, and also approximate the genotypic diversity in the population, usually with a test-function, such as a Gaussian shape, where the most dominant genotype (the mean) and the heterogeneity (the width) are well-defined. In standard coarse-graining macroscopic description of evolutionary game theory with competing species, such as in the study of cancer progression \cite{bukkuri2022life,cunningham2011evolutionary} and optimizing chemotherapeutic treatment \cite{stavnkova2019optimizing} via G-function \cite{vincent2005evolutionary,vincent1988evolution}, these two features (the mean and the width) are the most important mesoscopic variables. 

Perhaps even more interesting, the Rayleigh-Schr\"{o}dinger perturbation theory \cite{cohen1982rayleigh} can be utilized to answer a biological ``why'' question. Specifically, we have shown quantitative supporting evidence for why stress-induced mutagenesis exhibits a sharp transition rather than a gradual change in mutation rates, which is commonly observed in microbial and cancer cells \cite{cirz2005inhibition,bos2015emergence}. In contrast to physics, where phenomena may arise spontaneously, the complex emergent behavior observed in biology is the product of billions of years of natural selection, which has relentlessly honed and optimized the living systems we see today \cite{darwin2004origin,dawkins2016selfish}. It is therefore essential to focus on understanding why biological phenomena occur, rather than simply how they occur \cite{elsasser2016physical,dawkins1996blind}. Our findings highlight the significance of our approach as a valuable framework for modeling biological evolution, rather than just a mere mathematical exercise. 

 The methodology presented in this paper open up many avenues for future research. Although our study focused on a static ecological system, it is essential to acknowledge that most ecological systems in the real world are highly complex \cite{bhattacharjee2019bacterial,phan2020bacterial} and dynamic in nature \cite{fu2018spatial}. Therefore, one possible direction for future research is to extend our analogy to incorporate the effects of dynamical ecological systems, such as seasonal change or periodic cycle of drug-administration \cite{phan2021it}, in which it is expected that quantum mechanical methods to deal with temporal-varying Hamiltonian (e.g. time-dependent perturbation theory \cite{langhoff1972aspects}, adiabatic invariant \cite{dykhne1960quantum}, Floquet theory \cite{casas2001floquet}) can be employed. Another exciting adventure is to explore the impact of landscape topology. In particular, it would be interesting to investigate whether certain topological features of the fitness landscape can amplify or suppress the effects of stress-induced mutagenesis on population dynamics, since spatial topology has been shown to affect the quantization conditions greatly \cite{wang2015new}. Finally, there have been recent attempts to investigate exotic collective behaviors and new sectors of evolutionary dynamics using robots with engineered ecological interactions \cite{wang2021emergent,phan2021bootstrapped} and stress-induced mutable genomes \cite{wang2022robots}, which might allow us to see the realization of our analogy in a physical evolvable system beyond biology. So much to do; the future seems bright and exciting.

\section{Acknowledgement}

We thank Robert H. Austin, Kenneth J. Pienta, Joel Brown, Emma U. Hammarlund and Sarah R. Amend for the chance to give a talk on this simple but curious finding at Moffit Cancer Center (2021) and many useful discussions followed, which motivated us to share it with a wider audience.  We also thank Truong H. Cai and Ramzi Khuri for insightful comments.

\section{Declarations}

\begin{itemize}
\item Funding: This research received no external funding.
\item Conflict of interest: The authors declare no conflict of interest. 
\item Ethics approval: Not applicable. 
\item Consent to participate: Not applicable.
\item Consent for publication: Not applicable.
\item Availability of data and materials: Not applicable.
\item Code availability: The MatLab codes for the simulations used in this study are available from the corresponding author upon request. 
\item Authors' contributions: Conceptualization, T.V.P and D.K.T; analytical investigation, D.V.T., V.D.A. and K.T.P.; writing---original draft preparation, D.V.T., V.D.A., K.T.P., D.K.T. and T.V.P.; writing---review and editing, D.V.T., V.D.A., K.T.P., D.M.N., D.K.T. and T.V.P.; visualization, T.V.P.; supervision, H.D.T., V.H.D. and T.V.P.; project administration, T.V.P.. All authors provided critical feedback and helped shape the research, analysis and manuscript. All authors have read and agreed to the published version of the manuscript.
\end{itemize}

\appendix

\section{Summary of all Mathematical Quantities \label{summary_math}}

Here we summarize all quantities in our proposed model for the evolution dynamics with stress-induced mutagenesis:
\begin{itemize}
    \item $t$: time.
    \item $\vec{x}$: position (a genomic configuration) in the abstract $\mathscrsfs{D}$-dimensional fitness landscape.

    \item $b(\vec{x},t)$: population density on the landscape, which has the unit of population number per unit-volume (equal to a unit-length to the power $\mathscrsfs{D}$).

    \item $D(\vec{x})$: effective diffusivity in the landscape, which has the unit of unit-length squared (to the power $2$) per unit-time. 
    
    \item $R(\vec{x})$: the maximum growth-rate of the sub-population located at position $\vec{x}$ in the landscape, which has the unit of inverse unit-time.

    \item $K$: carrying capacity, which has the unit of population number.

    \item $S$: success, which is the ratio between the total population number $\int d^{\mathscrsfs{D}} \vec{x} b(\vec{x},t)$ and the carrying capacity $K$, therefore it is a dimensionless quantity. 
    
\end{itemize}

\section{Estimations of Stationary Population Success}

Let us define:
\begin{equation}
\Phi = \frac{1-S_{\text{st}}}{2D} R_0 \lambda^{-\gamma} \ ,
\end{equation}
so that the potential as in Eq. \eqref{identification} can be rewritten as:
\begin{equation}
V(S_{\text{st}},\vec{x}) = -\Phi \lambda^\gamma + \Phi |\vec{x}|^\gamma  \ .
\end{equation}
We want to estimate the ground state energy $E_{\Omega}$ of a pure power-law potential $U(\vec{x})=\Phi |\vec{x}|^\gamma$, which can be related to the ground state energy $\tilde{E}_{\Omega}$ of the potential $\tilde{U}(\vec{x}) = |\vec{x}|^\gamma$:
\begin{equation}
\tilde{E}_{\Omega} = \Phi^{\frac{2}{2+\gamma}} \tilde{E}_{\Omega} \ .
\end{equation}
If we can estimate $\tilde{E}_{\Omega}$, then we can estimate the stationary population success $S_{\text{st}}$ via the equality Eq. \eqref{set_to_0}:
\begin{equation}
-\Phi \lambda^\gamma + E_{\Omega} = 0 \ \ \Longrightarrow \ \ S_{\text{st}} = 1 - \frac{2D}{R_0 \lambda^2} \tilde{E}_{\Omega}^{\frac{2+\gamma}{\gamma}} \ .
\label{S_recover_dim}
\end{equation}
Note that, it is possible that the mathematical estimation of $S_{\text{st}}$ can become smaller than $0$ or larger than $1$, which is physically impossible. In that case, we can interpret these results as $S_{\text{st}}^{\text{(estimation)}} \rightarrow 0$ if $S_{\text{st}}^{\text{(estimation)}} < 0$ (population extinction), or $S_{\text{st}}^{\text{(estimation)}} \rightarrow 1$ if $S_{\text{st}}^{\text{(estimation)}} > 1$.

For simplicity, we will work with $\tilde{U}(\vec{x})$ instead of $U(\vec{x})$. The relationship between the ground states $\Psi_{\Omega}(\vec{x})$ and $\tilde{\Psi}_{\Omega}(\vec{x})$ are given by:
\begin{equation}
\Psi_{\Omega}(\vec{x}) = \Phi^{\frac{\mathscrsfs{D}}{2+\gamma}}  \tilde{\Psi}_\Omega \left( \Phi^{-\frac1{2+\gamma}} \vec{x}\right) \ .
\end{equation}

\subsection{Application of the Rayleigh-Ritz Variational Method \label{rayleigh_ritz}}

 For the Rayleigh-Ritz variational method \cite{rayleigh1907dynamical,ritz1909neue}, we need a trial-wavefunction. We choose a Gaussian function centered at $\vec{x}=0$ and has the standard deviation $\sigma$ as a parameter:
\begin{equation}
\tilde{\Psi}_{\text{trial}}(\sigma,\vec{x}) \propto \exp\left( - \frac{|\vec{x}|^2}{2\sigma^2}\right)  \ .
\end{equation}
A upper bound for $\tilde{E}_\Omega$ can be estimated via the following minimization with respect to $\sigma$:
\begin{equation}
\begin{split}
\tilde{E}_\Omega & \leq \min_{\sigma} \left[ \frac{ \int d^{\mathscrsfs{D}} \vec{x} \tilde{\Psi}_{\text{trial}} (\sigma,\vec{x}) \hat{\tilde{H}} \tilde{\Psi}_{\text{trial}} (\sigma,\vec{x}) }{ \int d^{\mathscrsfs{D}} \vec{x} \tilde{\Psi}_{\text{trial}} (\sigma,\vec{x}) \tilde{\Psi}_{\text{trial}} (\sigma,\vec{x}) }\right]
\\
&= \min_{\sigma} \left\{ \frac{ \int d^{\mathscrsfs{D}} \vec{x} \tilde{\Psi}_{\text{trial}} (\sigma,\vec{x}) \left[ -\frac12 \nabla^2 + \tilde{U}(\vec{x}) \right] \tilde{\Psi}_{\text{trial}} (\sigma,\vec{x}) }{ \int d^{\mathscrsfs{D}} \vec{x} \tilde{\Psi}^2_{\text{trial}} (\sigma,\vec{x}) } \right\} = \tilde{E}_\Omega^{\text{(RR)}} \ .
\end{split}
\end{equation}
Let us evaluate the function inside $\{ ...\}$:
\begin{equation}
\begin{split}
F(\sigma) = \frac{ \int^{\infty}_0 d|\vec{x}| |\vec{x}|^{\mathscrsfs{D}-1} 
 \exp\left( - \frac{|\vec{x}|^2}{2\sigma^2}\right) \left[ -\frac12 \nabla^2 + |\vec{x}|^\gamma \right] \exp\left( - \frac{|\vec{x}|^2}{2\sigma^2}\right) }{\int^{\infty}_0 d|\vec{x}| |\vec{x}|^{\mathscrsfs{D}-1} 
\exp\left( - \frac{|\vec{x}|^2}{\sigma^2}\right) }
\\
= \frac{ f_{\text{kin}}(\sigma) + f_{\text{pot}}(\sigma) }{\frac12 \Gamma \left( \frac{\mathscrsfs{D}}2 \right) \sigma^{\mathscrsfs{D}}} \ .
 \end{split} 
\end{equation}
This has two components, the potential part:
\begin{equation}
\begin{split}
f_{\text{pot}}(\sigma) &= \int^{\infty}_0 d|\vec{x}| |\vec{x}|^{\mathscrsfs{D}-1} 
 \exp\left( -\frac{|\vec{x}|^2}{2\sigma^2}\right) |\vec{x}|^\gamma  \exp\left( -\frac{|\vec{x}|^2}{2\sigma^2}\right)
 \\
 &= \int^{\infty}_0 d|\vec{x}| |\vec{x}|^{\mathscrsfs{D}+\gamma-1}
\exp\left( -\frac{|\vec{x}|^2}{\sigma^2}\right) = \frac12 \Gamma\left(\frac{\mathscrsfs{D}+\gamma}{2} \right) \sigma^{\mathscrsfs{D}+\gamma} \ ,
 \end{split}
\end{equation}
and the kinetic part:
\begin{equation}
\begin{split}
f_{\text{kin}}(\sigma) &= -\frac12 \int^{\infty}_0 d|\vec{x}| |\vec{x}|^{\mathscrsfs{D}-1} 
 \exp\left( -\frac{|\vec{x}|^2}{2\sigma^2}\right) \nabla^2  \exp\left( -\frac{|\vec{x}|^2}{2\sigma^2}\right)
 \\
 &= -\frac12 \int^{\infty}_0 d|\vec{x}| |\vec{x}|^{\mathscrsfs{D}-1}
 \exp\left( -\frac{|\vec{x}|^2}{\sigma^2}\right)  \left( \frac{|\vec{x}|^2 - \mathscrsfs{D} \sigma^2 }{\sigma^4} \right) = \frac14 \Gamma\left(\frac{\mathscrsfs{D}+2}2 \right) \sigma^{\mathscrsfs{D}-2}\ ,
 \end{split}
\end{equation}
where operating the Laplacian on any function $A(|\vec{x}|)$ gives:
\begin{equation}
\nabla^2 A(|\vec{x}|) = \frac1{|\vec{x}|^{\mathscrsfs{D}-1}} \partial_{|\vec{x}|} \left[ |\vec{x}|^{\mathscrsfs{D}-1} \partial_{|\vec{x}|} A(|\vec{x}|) \right] \ .
\end{equation}
Together these two contributions, we have:
\begin{equation}
F(\sigma) = \frac{ \frac14 \Gamma\left(\frac{\mathscrsfs{D}+2}2 \right) \sigma^{\mathscrsfs{D}-2} + \frac12 \Gamma\left(\frac{\mathscrsfs{D}+\gamma}{2} \right) \sigma^{\mathscrsfs{D}+\gamma} }{\frac12 \Gamma\left( \frac{\mathscrsfs{D}}2 \right) \sigma^{\mathscrsfs{D}}} = \frac{\mathscrsfs{D}}{4} \sigma^{-2} + \frac{\Gamma\left( \frac{\mathscrsfs{D}+\gamma}2\right)}{\Gamma\left( \frac{\mathscrsfs{D}}2\right)} \sigma^\gamma \ .
\end{equation}
The minimum of $F(\sigma)$ can be found by using the Cauchy inequality:
\begin{equation}
\begin{split}
F(\sigma) & = \gamma \left( \frac{\mathscrsfs{D}}{4\gamma} \sigma^{-2} \right) + 2\left[ \frac{\Gamma\left( \frac{\mathscrsfs{D}+\gamma}2\right)}{2\Gamma\left( \frac{\mathscrsfs{D}}2\right)} \sigma^\gamma \right] 
\\
& \geq (2+\gamma)\left\{ \left( \frac{\mathscrsfs{D}}{4\gamma} \right)^{\gamma} \left[ \frac{\Gamma\left( \frac{\mathscrsfs{D}+\gamma}2\right)}{2\Gamma\left( \frac{\mathscrsfs{D}}2\right)}  \right]^2 \right\}^{\frac1{2+\gamma}} = F(\sigma_{\min})\ ,
\end{split}
\end{equation}
the equal sign appears when:
\begin{equation}
\frac{\mathscrsfs{D}}{4\gamma} \sigma_{\min}^{-2} = \frac{\Gamma\left( \frac{\mathscrsfs{D}+\gamma}2\right)}{2\Gamma\left( \frac{\mathscrsfs{D}}2\right)} \sigma_{\min}^\gamma \ \ \Longrightarrow \ \ \sigma_{\min} = \left[ \frac{\mathscrsfs{D}}{2\gamma} \frac{\Gamma\left( \frac{\mathscrsfs{D}}2\right)}{\Gamma\left( \frac{\mathscrsfs{D}+\gamma}2\right)} \right]^{\frac1{2+\gamma}} \ .
\label{sigma_min}
\end{equation}
Hence we get an upper estimation for $\tilde{E}_{\Omega}$, i.e. $\tilde{E}_{\Omega} \leq F(\sigma_{\min})=\tilde{E}_{\Omega}^{\text{(RR)}}$. Plug this inside Eq. \eqref{S_recover_dim}, we can have a lower-bound estimate the stationary population success:
\begin{equation}
\begin{split}
S_{\text{st}} & \geq S^{\text{(RR)}}_{\text{st}} = 1 - \frac{2D}{R_0 \lambda^2} \left[\tilde{E}^{\text{(RR)}}_\Omega\right]^{\frac{2+\gamma}{\gamma}}
\\
&= 1 - \frac{2D}{R_0 \lambda^2}  (2+\gamma)^{\frac{2+\gamma}{\gamma}} \left( \frac{\mathscrsfs{D}}{4\gamma} \right) \left[ \frac{\Gamma\left( \frac{\mathscrsfs{D}+\gamma}2\right)}{2\Gamma\left( \frac{\mathscrsfs{D}}2\right)}  \right]^{\frac{2}{\gamma}} \ .
\end{split}
\label{Sst_RR}
\end{equation}

\subsection{Application of the Weinstein Method \label{weinstein}}

Weinstein method \cite{weinstein1934modified,pollak2019tight} picks up where the Rayleigh-Ritz has left off, using the trial-wavefunction $\tilde{\Psi}(\sigma_{\min},\vec{x})$ to estimate the lower-bound of $\tilde{E}_{\Omega}$:
\begin{equation}
\tilde{E}_{\Omega} \geq {\tilde{E}_\Omega^{\text{(RR)}}} - \left\{ \langle \hat{\tilde{H}}^2\rangle_{\sigma_{\min}} - \left[ \tilde{E}_\Omega^{\text{(RR)}}\right]^2 \right\}^{\frac12} = \tilde{E}_{\Omega}^{\text{(W)}} \ ,
\label{Weinstein_formula}
\end{equation}
where $\sigma_{\min}$ is as found in Eq. \eqref{sigma_min} and:
\begin{equation}
\begin{split}
 \langle \hat{\tilde{H}}^2\rangle_{\sigma}   &= \frac{ \int d^{\mathscrsfs{D}} \vec{x} \tilde{\Psi}_{\text{trial}}(\sigma_{\min},\vec{x}) \hat{\tilde{H}}^2  \tilde{\Psi}_{\text{trial}} (\sigma,\vec{x}) }{ \int d^{\mathscrsfs{D}} \vec{x} \tilde{\Psi}^2_{\text{trial}} (\sigma,\vec{x}) } \Bigg|_{\sigma = \sigma_{\min}}
 \\
 & = \frac{ \int^{\infty}_0 d|\vec{x}| |\vec{x}|^{\mathscrsfs{D}-1} 
 \exp\left( - \frac{|\vec{x}|^2}{2\sigma^2}\right) \left[ -\frac12 \nabla^2 + |\vec{x}|^\gamma \right]^2 \exp\left( - \frac{|\vec{x}|^2}{2\sigma^2}\right) }{\frac12 \Gamma \left( \frac{\mathscrsfs{D}}2 \right) \sigma^{\mathscrsfs{D}}} \Bigg|_{\sigma = \sigma_{\min}} \ .
\end{split}
\end{equation}
The numerator of this expression can be divided into four parts, evaluated separately. Open up the operator $\left[ -\frac12 \nabla^2 + |\vec{x}|^\gamma \right]^2$, we get the contribution from the $|\vec{x}|^{\gamma}|\vec{x}|^{\gamma}$ term:
\begin{equation}
\begin{split}
&N_1(\sigma) =    \int^{\infty}_0 d|\vec{x}| |\vec{x}|^{\mathscrsfs{D}-1} 
 \exp\left( - \frac{|\vec{x}|^2}{2\sigma^2}\right) |\vec{x}|^{2\gamma} \exp\left( - \frac{|\vec{x}|^2}{2\sigma^2}\right) 
 \\
 & \ \ = \int^{\infty}_0 d|\vec{x}| |\vec{x}|^{\mathscrsfs{D}+2\gamma-1} 
 \exp\left( - \frac{|\vec{x}|^2}{\sigma^2}\right) = \frac12 \Gamma\left( \frac{\mathscrsfs{D}+2\gamma}{2} \right) \sigma^{\mathscrsfs{D}+2\gamma} \ ,
\end{split}
\end{equation}
the $-\frac12 |\vec{x}|^\gamma \nabla^2$ term:
\begin{equation}
\begin{split}
N_2(\sigma) &= -\frac12   \int^{\infty}_0 d|\vec{x}| |\vec{x}|^{\mathscrsfs{D}-1} 
 \exp\left( - \frac{|\vec{x}|^2}{2\sigma^2}\right) |\vec{x}|^{\gamma} \nabla^2  \exp\left( - \frac{|\vec{x}|^2} {2\sigma^2}\right)  
 \\
 & = -\frac12 \int^{\infty}_0 d|\vec{x}| |\vec{x}|^{\mathscrsfs{D}+\gamma-1}
 \exp\left( -\frac{|\vec{x}|^2}{\sigma^2}\right)  \left(\frac{|\vec{x}|^2 - \mathscrsfs{D} \sigma^2 }{\sigma^4} \right)  
 \\
 & \ \ = \frac{1}{8}\left(\mathscrsfs{D}-\gamma \right) \Gamma\left(\frac{\mathscrsfs{D}+\gamma}{2}\right) \sigma^{\mathscrsfs{D}+\gamma-2} \ ,
\end{split}
\end{equation}
the $-\frac12 \nabla^2 |\vec{x}|^\gamma$ term:
\begin{equation}
\begin{split}
N_3(\sigma) &= -\frac12   \int^{\infty}_0 d|\vec{x}| |\vec{x}|^{\mathscrsfs{D}-1} 
 \exp\left( - \frac{|\vec{x}|^2}{2\sigma^2}\right) \nabla^2 \left[  |\vec{x}|^{\gamma} \exp\left( - \frac{|\vec{x}|^2} {2\sigma^2}\right) \right] 
 \\
 & = -\frac12   \int^{\infty}_0 d|\vec{x}| |\vec{x}|^{\mathscrsfs{D} + \gamma-3}
 \exp\left( - \frac{|\vec{x}|^2}{\sigma^2}\right) 
 \\
 & \ \ \ \ \ \ \ \ \ \ \ \ \ \ \ \ \ \ \ \ \ \ \ \ \ \ \ \left[\frac{|\vec{x}|^4 - (\mathscrsfs{D}+2\gamma) \sigma^2 |\vec{x}|^2  + \gamma(\mathscrsfs{D}+\gamma-2) \sigma^4}{\sigma^4} \right]
 \\
 &= \frac{1}{8}\left(\mathscrsfs{D}-\gamma \right) \Gamma\left(\frac{\mathscrsfs{D}+\gamma}{2}\right) \sigma^{\mathscrsfs{D}+\gamma-2} \ ,
\end{split}
\end{equation}
and the $\frac14 \nabla^2 \nabla^2$ term:
\begin{equation}
\begin{split}
N_4(\sigma) &=\frac14    \int^{\infty}_0 d|\vec{x}| |\vec{x}|^{\mathscrsfs{D}-1} 
 \exp\left( - \frac{|\vec{x}|^2}{2\sigma^2}\right) \nabla^2 \left[ \nabla^2 \exp\left( - \frac{|\vec{x}|^2}{2\sigma^2}\right) \right]
 \\
 & = \frac14    \int^{\infty}_0 d|\vec{x}| |\vec{x}|^{\mathscrsfs{D}-1}  \exp\left( - \frac{|\vec{x}|^2}{\sigma^2}\right) \left[ \frac{ |\vec{x}|^4 - 2(\mathscrsfs{D}+2)\sigma^2 |\vec{x}|^2 + \mathscrsfs{D}(\mathscrsfs{D}+2)\sigma^4 }{\sigma^8} \right]  
 \\ 
 & = \frac1{32} \mathscrsfs{D}(\mathscrsfs{D}+2) \Gamma\left(\frac{\mathscrsfs{D}}{2} \right) \sigma^{\mathscrsfs{D}-4} \ .
\end{split}
\end{equation}
Thus, follow from Eq. \eqref{Weinstein_formula}, we obtain:
\begin{equation}
\tilde{E}_{\Omega}^{\text{(W)}} = {\tilde{E}_\Omega^{\text{(RR)}}} - \left\{ \frac{\sum^4_{j=1} N_j(\sigma)}{\frac12 \Gamma \left( \frac{\mathscrsfs{D}}2\right) \sigma^{\mathscrsfs{D}}} \Bigg|_{\sigma = \sigma_{\min}}- \left[\tilde{E}_\Omega^{\text{(RR)}}\right]^2 \right\}^{\frac12} \ ,
\end{equation}
in which we can estimate the upper-bound for the stationary population success as:
\begin{equation}
 S_{\text{st}} \leq S^{\text{(W)}}_{\text{st}} = 1 - \frac{2D}{R_0 \lambda^2} \left[\tilde{E}^{\text{(W)}}_\Omega\right]^{\frac{2+\gamma}{\gamma}}  
 \label{Sst_W}
\end{equation}

In Fig. \ref{fig2}, for the dependence of $S_{\text{st}}^{\text{(W}}$ on the exponent $\gamma$, we see a sharp turn at $\gamma=2$, where $\tilde{E}^{\text{(W)}}=\tilde{E}^{\text{(RR)}}$. To understand this, let us take a look at how the difference between them progresses with $\gamma$ by rewriting Eq. \eqref{Weinstein_formula} as follows:
\begin{equation}
\tilde{E}_{\text{st}}^{\text{(RR)}} - \tilde{E}_{\text{st}}^{\text{(W)}} = \left[ \langle \hat{H}^2 \rangle_\gamma - \langle \hat{H}\rangle_\gamma^2 \right]^{\frac12} \ ,
\end{equation}
in which we use:
\begin{equation}
\begin{split}
\langle \hat{H}^2 \rangle_\gamma &= \Big\langle \Psi_{\text{trial}}[\sigma_{\min}(\gamma)] \Big\vert \hat{H}^2 \Big\vert \Psi_{\text{trial}}[\sigma_{\min}(\gamma)] \Big\rangle \ ,
\\
\langle \hat{H} \rangle_\gamma^2 &= \Big\langle \Psi_{\text{trial}}[\sigma_{\min}(\gamma)] \Big\vert \hat{H} \Big\vert \Psi_{\text{trial}}[\sigma_{\min}(\gamma)] \Big\rangle^2 \ .
\end{split}
\end{equation}
Since $\langle \hat{H}^2 \rangle_\gamma$ and $\langle \hat{H}\rangle_\gamma^2$ are an analytical functions of $\gamma$, and mathematically we always have the inequality $\langle \hat{H}^2 \rangle_\gamma \geq \langle \hat{H}\rangle_\gamma^2$, thus in the limit $\gamma \rightarrow 2$ where the equal sign happens, the leading-order of the $(\gamma-2)$-expansion there should be as least 2nd-order:
\begin{equation}
\langle \hat{H}^2 \rangle_\gamma - \langle \hat{H}\rangle_\gamma^2 \propto (\gamma-2)^2 \ \ \Longrightarrow \ \ \tilde{E}_{\text{st}}^{\text{(RR)}} - \tilde{E}_{\text{st}}^{\text{(W)}} \propto |\gamma-2| \ ,
\end{equation}
therefore signals a sharp turn for $S^{\text{(W)}}_{\text{st}}$ at $\gamma=2$ due to the contribution from this non-differentiable absolute value function, as $S^{\text{(RR)}}_{\text{st}}$ is smooth there. This is indeed the behavior we have observed.

\subsection{Application of the Wentzel–Kramers–Brillouin Approximation  \label{wkb}}

The Wentzel–Kramers–Brillouin approximation in $\mathcal{D}=1$ use semiclassical quantization conditions to estimate the eigenenergies via a summation series \cite{vranivcar2000accuracy}:
\begin{equation}
\sum^{\infty}_{k=0} (-i)^{2k} \oint \Theta_{2k} = \sum^{\infty}_{k=0} = 2\pi \left( n+\frac12 \right) \ \ \text{where} \ \ n = \{ 0,1,2,3, ... \} \ .
\label{quant_cont}
\end{equation}
For the ground state, we consider $n=0$. 

Perhaps the most familiar part of this general formula is the 0th-order term, which is given by the total action in a single cycle of a classically-allowed periodic-trajectory:
\begin{equation}
\oint \Theta_0 = \oint dx \left\{ 2\left[ \tilde{E} - \tilde{U}(x) \right] \right\}^{\frac12} \ ,
\end{equation}
which can be evaluated with $\tilde{U}(x) = |x|^\gamma$ to be:
\begin{equation}
\begin{split}
\oint \Theta_0 = 4\sqrt{2} \int^{+\tilde{E}^{\frac1\gamma}}_{0} dx \left[\tilde{E} -|x|^{\gamma} \right]^{\frac12}
\ \xrightarrow{\ \ \rho=xE^{-\frac1\gamma} \ \ } \ 4\sqrt{2} \tilde{E}^{\frac{2+\gamma}{2\gamma}} \int^{+1}_{0} d\rho \left( 1-|\rho|^\gamma \right)^{\frac12} &
\\
 = 4\sqrt{2} \tilde{E}^{\frac{2+\gamma}{2\gamma}} \left[ \frac{\sqrt{\pi}}2 \frac{\Gamma\left( 1+\frac1\gamma \right)}{\Gamma\left( \frac32+\frac1\gamma \right)} \right] = \sqrt{8\pi} \frac{\Gamma\left( 1+\frac1\gamma \right)}{\Gamma\left( \frac32+\frac1\gamma \right)} \tilde{E}^{\frac{2+\gamma}{\gamma}} & \ .
\end{split}
\end{equation}
Using this in Eq. \eqref{quant_cont}, we can estimate the energy of the ground state:
\begin{equation}
\begin{split}
\sqrt{8\pi} \frac{\Gamma\left( 1+\frac1\gamma \right)}{\Gamma\left( \frac32+\frac1\gamma \right)} \left[ \tilde{E}_\Omega^{\text{(WKB,0)}}\right]^{\frac{2+\gamma}{\gamma}} = 2\pi \left( n+\frac12 \right)\Bigg|_{n=0} = \pi &
\\
\Longrightarrow \ \ \tilde{E}_\Omega^{\text{(WKB,0)}} = \left[ \sqrt{\frac{\pi}{8}} \frac{\Gamma\left( \frac32+\frac1\gamma \right)}{\Gamma\left( 1+\frac1\gamma \right)} \right]^{\frac{2\gamma}{2+\gamma}} & \ .
\end{split}
\end{equation}
We can take this result and apply Eq. \eqref{S_recover_dim} to get a 0th-order estimation for the stationary population success:
\begin{equation}
S^{\text{(WKB,0)}}_{\text{st}}  = 1 - \frac{2D}{R_0 \lambda^2} \left[\tilde{E}^{\text{(WKB,0)}}_\Omega\right]^{\frac{2+\gamma}{\gamma}} = 1-\frac{\pi}{4} \frac{D}{R_0 \lambda^2} \left[ \frac{\Gamma\left( \frac32+\frac1\gamma \right)}{\Gamma\left( 1+\frac1\gamma \right)} \right]^2 \ .
\label{Sst_WKB0}
\end{equation}

For higher-orders, the quantization conditions can be written as:
\begin{equation}
\sum^{\infty}_{k=0} c_{2k} \mathcal{E}^{\frac12 - k} = 2\pi \left( n+\frac12 \right) \ \ \text{where} \ \ \mathcal{E}=\tilde{E}^{\frac{2+\gamma}{\gamma}} \ .
\label{all_order}
\end{equation}
The first few coefficients we have calculated for the pure power-law potential $\tilde{U}(x)=|x|^\gamma$ to be:
\begin{equation}
\begin{split}
c_0 = 2\sqrt{2\pi} \frac{\Gamma\left(1+\frac1\gamma \right)}{\Gamma\left(\frac32 + \frac1\gamma \right)} \ , \ c_2 = -\frac{\sqrt{2\pi}}{12} \frac{\gamma \Gamma\left( 2 - \frac1\gamma\right)}{\Gamma\left( \frac12 - \frac1\gamma \right)} & \ ,
\\
c_4 = -\frac{\sqrt{2\pi}}{8640} \frac{\gamma^3 (3+2\gamma) \Gamma\left( 4-\frac{3}{\gamma}\right)}{(3-2\gamma) \Gamma\left( -\frac12 - \frac3\gamma \right)} & \ .
\end{split}
\end{equation}
For tractability, we will only go up the 2nd-order. Then, for the ground state, from Eq. \eqref{all_order} we can arrive at:
\begin{equation}
c_0 \mathcal{E}_\Omega^{\frac12} + c_2 \mathcal{E}_\Omega^{-\frac12} = 2\pi \left( n+\frac12 \right)\Bigg|_{n=0} = \pi \ ,
\end{equation}
which can be solved as a quadratic polynomial:
\begin{equation}
\mathcal{E}_{\Omega} = \frac{\pi^2-2c_0c_2 + \pi \sqrt{\pi^2 - 4c_0c_2}}{2c_0^2} \ ,
\end{equation}
hence by using Eq. \eqref{S_recover_dim} we can obtain the 2nd-order estimation:
\begin{equation}
S^{\text{(WKB,2)}}_{\text{st}} = 1-\frac{2D}{R_0 \lambda^2} \left( \frac{\pi^2-2c_0c_2 + \pi \sqrt{\pi^2 - 4c_0c_2}}{2c_0^2} \right) \ .
\label{Sst_WKB2}
\end{equation}

\section{Simulation of the Non-Homogeneous Random-Walk on the Landscape \label{simulation_explain}}

\begin{figure*}[ht]
\centering
\includegraphics[width=\textwidth,keepaspectratio]{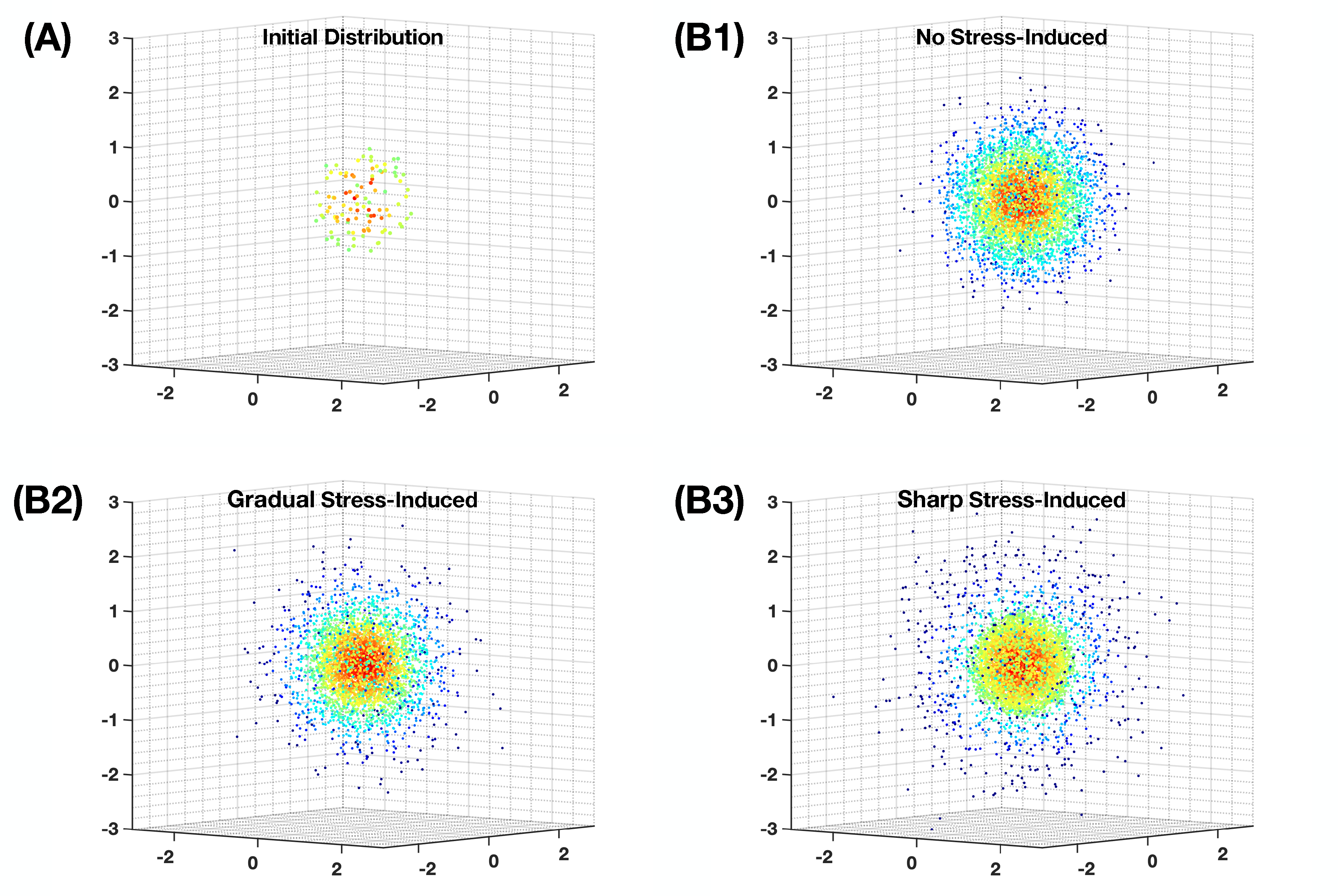}
\caption{\textbf{Our simulation for the evolution of heterogeneous population distribution on the $\bd=3$-dimensional fitness landscape as described in Eq. \eqref{toy_fitness}.} We use the parameter values $D=1/18$, $R_0=1$, $\lambda=1$, and $K=10^5$. For better visualization, only $10\%$ of the agents are shown. (A) The initial distribution at $t=0$ we use for all runs, using $\int d^3\vec{x} b(\vec{x},0)=10^3$ agents.  (B1) A snapshot of the distribution at stationary state if the evolution has no stress-induced mutagenesis. (B2) A snapshot of the distribution at stationary state if the evolution has gradual stress-induced mutagenesis as in Eq.\eqref{gradual_diff_rewrite}, where $\epsilon=0.1$. (B3) A snapshot of the distribution at stationary state if the evolution has has sharp stress-induced mutagenesis as in Eq.\eqref{sharp_diff_rewrite}, where $\epsilon=10$.}
\label{fig4}
\end{figure*}

We use an agent-based simulations to investigate the population dynamics, in which each agent are specified by its location $\vec{x}=\left( x_1,x_2,...,x_{\bd}\right)$ on the $\bd$-dimensional landscape. We discretize the time $t$ into evenly-pacing simulation time-steps, so that two consecutive steps are $\Delta t$ apart. At every simulation step, the position of each agents in every different direction $j\in \{ 1,2,3,...,\bd\}$ is updated with:
\begin{equation}
x_j (t+\Delta t) = x_j(t) + \left\{ 2 D\left[R(\vec{x})\right] \Delta t \times \mathcal{N}(0,1) \right\} \ , 
\end{equation}
where $D[R]$ is the fitness $R(\vec{x})$-dependence diffusivity, and $\mathcal{N}(0,1)$ is a sample values from a Gaussian distribution of mean value $0$ and standard deviation $1$. Each agent also have a chance to multiple (from one agent becomes two) or die. These two are controlled by a single value $p$:
\begin{equation}
p = R(\vec{x}) \left[1 - \frac{N(t)}{K} \right] \Delta t \ ,
\end{equation}
in which $K$ is the carrying capacity and $N(t)$ is the total number of agents at physical time $t$. A random number is generated uniformly between $[0,1]$, and if that number is larger than $|p|$ then the agent will multiply if $p>0$ and will die if $p<0$ (else, nothing will happen). 

For the fitness function $R(\vec{x})$, unless further specified, we use Eq. \eqref{toy_fitness}. For the gradual stress-induced mutagenesis regime, we use $D[R]$ as in Eq. \eqref{gradual_diff_rewrite}. For the sharp stress-induced mutagenesis regime, we use $D[R]$ as in Eq. \eqref{sharp_diff_rewrite}. At $t=0$, we use the very same initial distribution of $10^3$ agents in the landscape, which we place randomly using a uniformly-generated-procedure inside a ball of radius $\lambda$ (the \textit{fit region} on the landscape). 

For our simulations, we use a time-discretization $\Delta = 0.01$ and the total time of $T=100$. In every simulation, the population reach stationary state before $t=50$, so all the population distribution densities are temporal-averaging of all agent positions data in $t\in [51,100]$. The value of other parameters ($\bd$, $D$, $R_0$, $\lambda$, $K$, $\epsilon$) in different simulations are mentioned in the figures that show their results.

\section{Perturbative Corrections}
 
\subsection{With perturbed Hamiltonian contains $|\vec{x}|^2$ \label{perturbative_calc_1}}
Following Eq. \eqref{perturbative_correction}, with perturbed Hamiltonian Eq. \eqref{gradual_perturb_Hamiltonian}, we can estimate the ground state energy-shift via brute-force integration as follows:
\begin{equation}
\begin{split}
      \delta E_{\Omega}^{(1)} &=\frac{\displaystyle \int_0^{\infty} \dd |\vec{x}| |\vec{x}|^{\bd-1}  e^{-\frac{1}{2}\omega |\vec{x}|^2} \cdot \frac{1}{2}\hat{p}^2\left(\frac{\vert \vec{x} \vert}{\lambda} \right)^2 \cdot  e^{-\frac{1}{2}\omega |\vec{x}|^2}}{\displaystyle \int_0^{\infty} \dd |\vec{x}| |\vec{x}|^{\bd-1} e^{- \frac{1}{2}\omega |\vec{x}|^2} \cdot e^{- \frac{1}{2}\omega |\vec{x}|^2}}
      \\
      & = \frac{\displaystyle \int_0^{\infty} \dd |\vec{x}| |\vec{x}|^{\bd-1}  e^{-\frac{1}{2}\omega |\vec{x}|^2} \cdot \frac{1}{2}(-\nabla^2) \left[ \left(\frac{\vert \vec{x} \vert}{\lambda} \right)^2   e^{-\frac{1}{2}\omega |\vec{x}|^2}\right]}{\displaystyle \int_0^{\infty} \dd |\vec{x}| |\vec{x}|^{\bd-1} e^{-\omega |\vec{x}|^2}}
      \\
      &= \frac{\displaystyle -\frac12 \int_0^{\infty} \dd |\vec{x}| |\vec{x}|^{\bd-1}  e^{-\omega |\vec{x}|^2} \left[ \frac{\omega^2 |\vec{x}|^4 - \left( \mathscrsfs{D}+4\right)\omega |\vec{x}|^2 + 2\mathscrsfs{D}}{\lambda^2}\right]}{\displaystyle \frac12 \omega^{-\frac{\mathscrsfs{D}}{2}} \Gamma\left(\frac{\mathscrsfs{D}}{2} \right)} 
      \\
      &= \frac{\displaystyle -\frac{\left(\mathscrsfs{D}-2\right)\omega^{-\frac{\mathscrsfs{D}}{2}} \Gamma\left( 1+\frac{\mathscrsfs{D}}{2}\right)}{8\lambda^2}}{\displaystyle \frac12 \omega^{-\frac{\mathscrsfs{D}}{2}} \Gamma\left(\frac{\mathscrsfs{D}}{2} \right)} = \frac{\displaystyle \mathscrsfs{D} \left( \mathscrsfs{D} - 2 \right)}{\displaystyle 8\lambda^2}
      \ .
\end{split}
\label{app1}
\end{equation}

%%%%%%%%%%%%%%%%%%%%%%%
%%%%%%%%%%%%%%%%%%%%%%% 

\subsection{With perturbed Hamiltonian contains $x^\kappa$ \label{perturbative_calc_2}}

In order to evaluate $\delta E_\Omega^{(1)}$ for the $\hat{p}^2 \vert \Vec{x}\vert^\kappa$ perturbation operator, we start from Eq. \eqref{perturbative_correction}:
\begin{equation}
\begin{split}
      \delta E_{\Omega}^{(1)} &=\frac{\displaystyle \int_0^{\infty} \dd |\vec{x}| |\vec{x}|^{\bd-1}  e^{-\frac{1}{2}\omega |\vec{x}|^2} \cdot \frac{1}{2}\hat{p}^2\left(\frac{\vert \vec{x} \vert}{\lambda} \right)^\kappa \cdot  e^{-\frac{1}{2}\omega |\vec{x}|^2}}{\displaystyle \int_0^{\infty} \dd |\vec{x}| |\vec{x}|^{\bd-1} e^{- \frac{1}{2}\omega |\vec{x}|^2} \cdot e^{- \frac{1}{2}\omega |\vec{x}|^2}}
      \\
      & = \frac{\displaystyle \int_0^{\infty} \dd |\vec{x}| |\vec{x}|^{\bd-1}  e^{-\frac{1}{2}\omega |\vec{x}|^2} \cdot \frac{1}{2}(-\nabla^2) \left[ \left(\frac{\vert \vec{x} \vert}{\lambda} \right)^\kappa  e^{-\frac{1}{2}\omega |\vec{x}|^2}\right]}{\displaystyle \int_0^{\infty} \dd |\vec{x}| |\vec{x}|^{\bd-1} e^{-\omega |\vec{x}|^2}}
      \\
      &= \frac{\displaystyle -\frac12 \int_0^{\infty} \dd |\vec{x}| |\vec{x}|^{\bd+\kappa-1}  e^{-\omega |\vec{x}|^2} \left[ \frac{\omega^2 |\vec{x}|^4 - \left(\mathscrsfs{D}+2\kappa \right) \omega |\vec{x}|^2 + \kappa\left(\mathscrsfs{D}+\kappa-2\right)}{\lambda^\kappa}\right]}{\displaystyle \frac12 \omega^{-\frac{\mathscrsfs{D}}{2}} \Gamma\left(\frac{\mathscrsfs{D}}{2} \right)} 
      \\
      &= \frac{\displaystyle -\frac{\left(\mathscrsfs{D}-\kappa \right)\omega^{1-\frac{\mathscrsfs{D}+\kappa}{2}} \Gamma\left(\frac{\mathscrsfs{D}+\kappa}{2}\right)}{8\lambda^\kappa}}{\displaystyle \frac12 \omega^{-\frac{\mathscrsfs{D}}{2}} \Gamma\left(\frac{\mathscrsfs{D}}{2} \right)} = \frac{\displaystyle \left(\mathscrsfs{D} - \kappa \right) \omega^{1-\frac{\kappa}2}\Gamma\left(\frac{\mathscrsfs{D}+\kappa}2\right)}{\displaystyle 4 \lambda^\kappa \Gamma\left( \frac{\mathscrsfs{D}}2\right)}
      \ .
\end{split}
\end{equation}
For the sanity check, when we substitute $\kappa = 2$, the result $ \delta E_{\Omega}^{(1)}$ becomes Eq. \eqref{app1}.

%%%%%%%%%%%%%%%%%%%%%%%%
%%%%%%%%%%%%%%%%%%%%%%%%

\subsection{With perturbed Hamiltonian contains Heaviside function \label{perturbative_calc_3}}

In order to evaluate $\delta E_\Omega^{(1)}$ for the non-Hermitian operator containing Heaviside function, which is given in Eq. \eqref{Sharp_perturb_Hamiltonian}, we would like to expand Eq. \eqref{perturbative_correction}:
\begin{equation}
\begin{split}
      \delta E_{\Omega}^{(1)} &=\frac{\displaystyle \int_0^{\infty} \dd |\vec{x}| |\vec{x}|^{\bd-1}  e^{-\frac{1}{2}\omega |\vec{x}|^2} \cdot \frac{1}{2}\hat{p}^2 \Theta \left(\frac{\vert \vec{x} \vert}{\lambda} - 1 \right) \cdot  e^{-\frac{1}{2}\omega |\vec{x}|^2}}{\displaystyle \int_0^{\infty} \dd |\vec{x}| |\vec{x}|^{\bd-1} e^{- \frac{1}{2}\omega |\vec{x}|^2} \cdot e^{- \frac{1}{2}\omega |\vec{x}|^2}}
      \\
      & = \frac{\displaystyle \int_0^{\infty} \dd |\vec{x}| |\vec{x}|^{\bd-1}  e^{-\frac{1}{2}\omega |\vec{x}|^2} \cdot \frac{1}{2}(-\nabla^2) \left[ \Theta \left(\frac{\vert \vec{x} \vert}{\lambda} - 1 \right)  e^{-\frac{1}{2}\omega |\vec{x}|^2}\right]}{\displaystyle \int_0^{\infty} \dd |\vec{x}| |\vec{x}|^{\bd-1} e^{-\omega |\vec{x}|^2}}
      \\
      &= \frac{\displaystyle -\frac12 \int_0^{\infty} \dd |\vec{x}|  e^{-\frac{1}{2}\omega |\vec{x}|^2} \cdot \partial_{|\vec{x}|} \left\{|\vec{x}|^{\mathscrsfs{D}-1}\partial_{|\vec{x}|} \left[ \Theta \left(\frac{\vert \vec{x} \vert}{\lambda} - 1 \right)  e^{-\frac{1}{2}\omega |\vec{x}|^2}\right] \right\}}{\displaystyle \frac12 \omega^{-\frac{\mathscrsfs{D}}{2}} \Gamma\left(\frac{\mathscrsfs{D}}{2} \right)} \ .
\end{split}
\label{perturb_theta}
\end{equation}
It is non-trivial to deal with the derivatives of Heaviside function, so let us move slower:
\begin{equation}
\begin{split}
& \partial_{|\vec{x}|} \left\{|\vec{x}|^{\mathscrsfs{D}-1}\partial_{|\vec{x}|} \left[ \Theta \left(\frac{\vert \vec{x} \vert}{\lambda} - 1 \right)  e^{-\frac{1}{2}\omega |\vec{x}|^2}\right] \right\}
\\
& \ \ \ \ = \partial_{|\vec{x}|} \left\{|\vec{x}|^{\mathscrsfs{D}-1} e^{-\frac{1}{2}\omega |\vec{x}|^2} \left[-\omega |\vec{x}| \Theta \left(\frac{\vert \vec{x} \vert}{\lambda} - 1 \right) + \frac{1}{\lambda} \delta \left(\frac{\vert \vec{x} \vert}{\lambda} - 1 \right)\right] \right\} 
\\
& \ \ \ \ =\omega \left( \omega |\vec{x}|^2 - \bd \right) x^{\bd-1} e^{-\frac12 \omega |\vec{x}|^2} \Theta \left(\frac{\vert \vec{x} \vert}{\lambda} - 1 \right)
\\
& \ \ \ \ \ \ + \frac1{\lambda} \left[-2\omega |\vec{x}|^{\bd} + \left( \bd -1 \right)|\vec{x}|^{\bd -2} \right] e^{-\frac{1}{2}\omega |\vec{x}|^2} \delta \left(\frac{\vert \vec{x} \vert}{\lambda} - 1 \right)
\\
& \ \ \ \ \ \ + \frac1{\lambda} |\vec{x}|^{\mathscrsfs{D}-1} e^{-\frac{1}{2}\omega |\vec{x}|^2} \partial_{|\vec{x}|} \delta \left(\frac{\vert \vec{x} \vert}{\lambda} - 1 \right)
\ ,
\end{split}
\end{equation}
where $\delta(...)$ is the Dirac-delta function \cite{balakrishnan2003all}. To proceed, we evaluate the integration of each terms separately. The $\Theta$-term:
\begin{equation}
\begin{split}
& -\frac12 \int^\infty_0 d|\vec{x}| e^{-\frac12\omega|\vec{x}|^2} \left[ \omega \left( \omega |\vec{x}|^2 - \bd \right) x^{\bd-1} e^{-\frac12 \omega |\vec{x}|^2} \Theta \left(\frac{\vert \vec{x} \vert}{\lambda} - 1 \right) \right]
\\
& \ \ \ \ = -\frac12 \int^{\infty}_{\lambda} d|\vec{x}| x^{\bd-1} e^{-\omega|\vec{x}|^2} \omega \left( \omega |\vec{x}|^2 - \bd \right) 
\\
& \ \ \ \ = \frac14 \omega \lambda^{\bd} \left[-2 e^{-\omega \lambda^2} +  \omega \lambda^2 \expint_{-\frac{\bd}{2}}\left(\omega \lambda^2\right) \right] \ ,
\end{split}
\end{equation}
the $\delta$-term:
\begin{equation}
\begin{split}
&-\frac12 \int^\infty_0 d|\vec{x}| e^{-\frac12\omega|\vec{x}|^2}
\left\{ \frac1{\lambda} \left[-2\omega |\vec{x}|^{\bd} + \left( \bd -1 \right)|\vec{x}|^{\bd -2}  \right] e^{-\frac{1}{2}\omega |\vec{x}|^2} \delta \left(\frac{\vert \vec{x} \vert}{\lambda} - 1 \right) \right\}
\\ 
& \ \ \ \ = -\frac1{2\lambda} \left[-2\omega |\vec{x}|^{\bd} + \left( \bd -1 \right)|\vec{x}|^{\bd -2}  \right] e^{-\omega |\vec{x}|^2} \Bigg|_{|\vec{x}|=\lambda}
\\
& \ \ \ \ = -\frac1{2\lambda} \left[-2\omega \lambda^{\bd} + \left( \bd-1 \right)\lambda^{\bd -2}  \right] e^{-\omega \lambda^2} \ ,
\end{split}
\end{equation}
and finally the $\partial_{|\vec{x}|} \delta$-term:
\begin{equation}
\begin{split}
&-\frac12 \int^\infty_0 d|\vec{x}| e^{-\frac12\omega|\vec{x}|^2} \left[ \frac1{\lambda} |\vec{x}|^{\mathscrsfs{D}-1} e^{-\frac{1}{2}\omega |\vec{x}|^2} \partial_{|\vec{x}|} \delta \left(\frac{\vert \vec{x} \vert}{\lambda} - 1 \right) \right]
\\
& \ \ \ \ = -\frac1{2\lambda} \int^\infty_0 d|\vec{x}| |\vec{x}|^{\mathscrsfs{D}-1} e^{-\omega |\vec{x}|^2} \partial_{|\vec{x}|} \delta \left(\frac{\vert \vec{x} \vert}{\lambda} - 1 \right) 
\\
& \ \ \ \ = \frac1{2\lambda} \int^\infty_0 d|\vec{x}| \partial_{|\vec{x}|} \left( |\vec{x}|^{\mathscrsfs{D}-1} e^{-\omega |\vec{x}|^2} \right) \delta \left(\frac{\vert \vec{x} \vert}{\lambda} - 1 \right) 
\\
& \ \ \ \ = \frac1{2\lambda} |\vec{x}|^{\bd-2} \left( \bd - 2 \omega |\vec{x}|^2 - 1 \right) e^{-\omega |\vec{x}|^2} \Bigg|_{|\vec{x}|=\lambda}
\\
& \ \ \ \ = \frac1{2\lambda} \left[ - 2 \omega \lambda^\bd + \left( \bd - 1 \right) \lambda^{\bd-2}  \right] e^{-\omega \lambda^2} \ ,
\end{split}
\end{equation}
in which we have used integration-by-part. Adding up these three, we get the numerator of Eq. \eqref{perturb_theta}, thus gives:
\begin{equation}
\begin{split}
\delta E^{(1)}_{\Omega} &= \frac{\frac14 \omega \lambda^{\bd} \left[-2 e^{-\omega \lambda^2} +  \omega \lambda^2 \expint_{-\frac{\bd}{2}}\left(\omega \lambda^2\right) \right]}{\displaystyle \frac12 \omega^{-\frac{\mathscrsfs{D}}{2}} \Gamma\left(\frac{\mathscrsfs{D}}{2} \right)}
\\
& = \frac1{\lambda^2} \dfrac{\displaystyle(\omega \lambda^2)^{\frac{\bd}{2}+1}\left[-2 e^{-\omega \lambda^2}+\omega \lambda^2 \expint_{-\frac{\bd}{2}}\left(\omega \lambda^2\right)\right]}{\displaystyle2 \Gamma\left(\frac{\bd}{2}\right)} \ .
\end{split}
\end{equation}

\phantomsection
\bibliography{main}
\bibliographystyle{unsrt}

\end{document}